\documentclass{emulateapj}\usepackage{apjfonts}
\usepackage{graphicx}
\usepackage{CJKutf8}

\newcommand\p{\phantom{1}}

\newcommand\pgc{\ensuremath{p_{\rm GC}}}

\newcommand\etal{{et~al.}} 

\newcommand\feh{\ensuremath{[\hbox{Fe/H}]}}
\newcommand\Iacs{\ensuremath{I_{814}}}

\newcommand\gacs{\ensuremath{g_{475}}}
\newcommand\Vacs{\ensuremath{V_{606}}}
\newcommand\Hwfc{\ensuremath{H_{160}}}
\newcommand\zacs{\ensuremath{z_{850}}}

\newcommand{\vi}{\ensuremath{(V{-}I)}}

\newcommand{\viacs}{\ensuremath{V_{606}{-}I_{814}}}
\newcommand{\giacs}{\ensuremath{g_{475}{-}I_{814}}}
\newcommand{\gzacs}{\ensuremath{g_{475}{-}z_{850}}}
\newcommand{\ihacs}{\ensuremath{I_{814}{-}H_{160}}}
\newcommand{\vhacs}{\ensuremath{V_{606}{-}H_{160}}}
\newcommand{\VIacs}{\ensuremath{V_{606}{-}I_{814}}}
\newcommand{\gIacs}{\ensuremath{g_{475}{-}I_{814}}}
\newcommand{\IHacs}{\ensuremath{I_{814}{-}H_{160}}}
\newcommand{\VHacs}{\ensuremath{V_{606}{-}H_{160}}}
\newcommand{\gHacs}{\ensuremath{g_{475}{-}H_{160}}}
\newcommand{\VI}{\ensuremath{V{-}I}}
\newcommand{\gI}{\ensuremath{g{-}I}}
\newcommand{\gz}{\ensuremath{g{-}z}}
\newcommand{\IH}{\ensuremath{I{-}H}}
\newcommand\magauto{{\sc mag\_auto}}
\newcommand\kurt{\hbox{\it kurt}}
\newcommand\hst{{\it HST}}

\newcommand\lta{\lesssim}
\newcommand\gta{\gtrsim}
\newcommand\cote{C{\^ o}t{\' e}}
\newcommand\jordan{Jord{\'a}n}

\def\tmk#1{\tablenotemark{#1}}

\shortauthors{{Blakeslee, Cho, Peng, Ferrarese, Jord\'an \& Martel}}
\shorttitle{NGC\,1399 Color-Metallicity Nonlinearity}
\slugcomment{To appear in ApJ, January 2012}

\begin{document}
\begin{CJK}{UTF8}{}

\title{Optical and IR Photometry of Globular Clusters in NGC\,1399:\\
  Evidence for Color-Metallicity Nonlinearity\altaffilmark{*}}

\altaffiltext{*}{Based on observations with the NASA/ESA \textit{Hubble Space
  Telescope}, obtained from the Space Telescope Science
  Institute, which is operated by AURA, Inc.,
  under NASA contract NAS 5-26555.}

\author{John P.~Blakeslee\altaffilmark{1}, Hyejeon Cho (\CJKfamily{mj}조혜전)\altaffilmark{2}, 
Eric W.~Peng\altaffilmark{3,4},
Laura~Ferrarese\altaffilmark{1},
Andr\'es Jord\'an\altaffilmark{5},
\&
Andr\'e R.~Martel\altaffilmark{1,6}
}

\altaffiltext{1}{Dominion Astrophysical Observatory, Herzberg Institute of Astrophysics, National Research
  Council of Canada, Victoria, BC V9E\,2E7, Canada; {John.Blakeslee@nrc.ca}}
\altaffiltext{2}{Department of Astronomy and Center for Galaxy Evolution Research, Yonsei University, Seoul 120-749, Korea}
\altaffiltext{3}{Department of Astronomy, Peking University, Beijing 100871, China}
\altaffiltext{4}{Kavli Institute for Astronomy and Astrophysics, Beijing 100871, China}
\altaffiltext{5}{Departamento de Astronom\'ia y Astrof\'isica,
Pontificia Universidad Cat\'olica de Chile, 7820436 Macul, Santiago, Chile}
\altaffiltext{6}{Space Telescope Science Institute, 3700 San Martin Drive, Baltimore, MD 21218, USA}

\begin{abstract}
We combine new Wide Field Camera~3 IR Channel (WFC3/IR) F160W (\Hwfc) imaging data for
NGC\,1399, the central galaxy in the Fornax cluster, with archival F475W (\gacs), F606W
(\Vacs), F814W (\Iacs), and F850LP (\zacs) optical data from the Advanced Camera for
Surveys (ACS).  The purely optical \gIacs, \VIacs, and \gzacs\ colors of NGC\,1399's
rich globular cluster (GC) system exhibit clear bimodality, at least for magnitudes
$\Iacs>21.5$.  The optical-IR \IHacs\ color distribution appears unimodal, and this
impression is confirmed by mixture modeling analysis.  The \VHacs\ colors show marginal
evidence for bimodality, consistent with bimodality in \VIacs\ and unimodality in \IHacs.
If bimodality is imposed for \IHacs\ with a double Gaussian model, the preferred blue/red
split differs from that for optical colors; these ``differing bimodalities'' mean that
the optical and optical-IR colors cannot both be linearly proportional to metallicity.
Consistent with the differing color distributions,
the dependence of \IHacs\ on \gIacs\ for the matched GC sample is significantly
nonlinear, with an inflection point near the trough in the \gIacs\ color distribution;
the result is similar for the \IHacs\ dependence on \gzacs\ colors taken from the ACS
Fornax Cluster Survey.  These \gzacs\ colors have been calibrated empirically against
metallicity; applying this calibration yields a continuous, skewed, but single-peaked
metallicity distribution.  Taken together, these results indicate that nonlinear
color-metallicity relations play an important role in shaping the observed bimodal
distributions of optical colors in extragalactic GC systems.
\end{abstract}


\section{Introduction}
\label{sec:intro}

The major star formation episodes in the history of any large galaxy will be
imprinted in the properties of the galaxy's star cluster population.  Interpreting the
observed properties to derive the formation histories has proven to be a
difficult task.  All giant elliptical galaxies contain large populations of globular
clusters (GCs), often numbering in the thousands (e.g., Harris 1991).  The GC systems
generally follow bimodal distributions in optical colors (Zepf \& Ashman 1993;
Gebhardt \& Kissler-Patig 1999).  For giant ellipticals in clusters, the two peaks in
the color distributions are roughly equal in size (e.g., Peng \etal\ 2006; Harris \etal\
2006), except at large radii where the color distribution is more strongly weighted 
towards the blue (e.g., Dirsch \etal\ 2003; Harris 2009).

If the optical colors are interpreted as a direct proxy for metallicity, then the
bimodality represents an extraordinary constraint on the star formation histories of
giant ellipticals.  In this case, the two color peaks represent two distinct cluster
populations, differing by roughly a factor of ten in mean metallicity.  The bulk of the
host galaxy's stellar mass would then likely originate from two distinct major formation
episodes.  Ashman \& Zepf (1992) originally predicted such GC bimodality based on the
idea that ellipticals form from gas-rich major mergers of late-type galaxies that already
possessed extensive metal-poor GC populations.  Because ellipticals are believed to have
formed in a more stochastic, hierarchical fashion, a number of other merger or accretion
scenarios were later proposed to account for the observed bimodality (Forbes
\etal\ 1997; Kissler-Patig \etal\ 1998b; C\^ot\'e \etal\ 1998; Beasley \etal\ 2002;
Kravtsov \& Gnedin 2005).  However, these different scenarios are often difficult to
distinguish observationally from one another, and none appears to account naturally for
all the data (see Peng \etal\ 2006).

More recently, the assumption of optical colors as a simple, linear proxy for
metallicity has been reexamined (Richtler 2006; Yoon \etal\ 2006).  It has been known
for many years that the slope of the metallicity as a function of optical color becomes shallower
(i.e., color becomes more sensitive to metallicity) at intermediate metallicities
(Kissler-Patig \etal\ 1998a).  Yoon \etal\ (2006) showed that the ``wavy'' nonlinear
color-metallicity relation predicted by their stellar population models matched, at
least qualitatively, the color-metallicity data assembled by Peng \etal\ (2006).  They
further pointed out that the ``projection'' from metallicity to color with such wavy
relations can produce bimodal color distributions from unimodal distributions in
metallicity.  Cantiello \& Blakeslee (2007) confirmed the Yoon \etal\ (2006) result, in
the sense that other sets of models that include realistic prescriptions for the 
behavior of the horizontal branch as a function of metallicity also give nonlinear
color-metallicity relations that can produce bimodal color distributions.  More
recently, Yoon et al.\ (2011a,b) find that the GC metallicity distributions
derived for several galaxies from optical colors using their model
color-metallicity distributions are not bimodal, but are similar to the
metallicity distributions found for stellar halos in elliptical galaxies.

This issue remains highly controversial, but it is clear that additional studies of 
GC metallicity distributions are needed.  In particular, it is important to constrain
independently the form of the metallicity distributions in galaxies with
prominently bimodal GC color distributions.  Large samples of spectroscopic metallicities
in extragalactic GC systems (more than a few percent of the total population) are now
becoming available (Beasley \etal\ 2008; Foster \etal\ 2010; Alves-Brito \etal\ 2011;
Caldwell \etal\ 2011; see Section~\ref{sec:sum} for a discussion of these results).  However,
at the distances of the Virgo and Fornax clusters, spectroscopic data of sufficient
quality remains observationally expensive, requiring multiple nights of ten-meter class
telescope time to amass significant samples.

Another way to constrain GC metallicity distributions is through a combination of
near-infrared (near-IR) and optical photometry (e.g., Kissler-Patig \etal\ 2002; Beasley
\etal\ 2002; Puzia \etal\ 2002; Kundu \& Zepf 2007; Kotulla \etal\ 2008; Chies-Santos et
al.\ 2011a,b).  Hybrid optical-IR colors such as $I{-}H$ or $I{-}K$ for old stellar
systems such as GCs are mainly determined by the mean temperature of stars on the red
giant branch, which in turn depends almost entirely on metallicity (e.g., Bergbusch \&
VandenBerg 2001; Yi \etal\ 2001; Dotter \etal\ 2007).  Optical-IR colors that involve
bluer passbands, such as $V{-}H$ or $B{-}K$, also have the strong metallicity dependence
from the giant branch temperature, but they have additional sensitivity to the
main sequence turnoff, which depends strongly on age, and to the horizontal branch
morphology, which behaves nonlinearly with metallicity and also depends on age
(e.g., Lee \etal\ 1994; Sarajedini \etal\ 1997; Dotter \etal\ 2010).  Although the
dependence of giant branch temperature (and thus of $I{-}H$ and similar colors) on
metallicity is not necessarily linear, there is no evidence for a sharply nonlinear
transition with metallicity, as occurs for the horizontal branch.

The extensive GC system of NGC\,1399, the dominant elliptical galaxy in the Fornax cluster, 
has been a frequent target for optical photometric and spectroscopic studies since the
pioneering work of Hanes \& Harris (1986).
At a distance of 20 Mpc (Blakeslee \etal\ 2009) Fornax is the second nearest galaxy
cluster after Virgo, and NGC\,1399 is at its dynamical center (Drinkwater et al.\ 2001).
NGC\,1399 was one of the first external galaxies reported as having a bimodal
metallicity distribution, based on optical colors (Ostrov \etal\ 1993).  Interestingly,
this galaxy provided the first indication of the importance of the detailed
shape of the color-metallicity relation: Kissler-Patig et al.\ (1998a) measured
spectroscopic metallicities for a sample of GCs in NGC\,1399 and found that the slope of
metallicity versus \vi\ color was significantly flatter than the extrapolation from
low-metallicity Galactic GCs.  NGC\,1399 also has the largest sample of measured GC
radial velocities to date (Schuberth et al.\ 2010).

Here we present new near-IR and optical photometry of GCs in NGC\,1399 using the Infrared
Channel of the 
Wide Field Camera~3 (WFC3/IR) and Wide Field Channel of the Advanced Camera for Surveys
(ACS/WFC) on board the \textit{Hubble Space Telescope} (\hst).  We take an empirical 
approach in this work, comparing the color distributions and examining the color-color
relations, without trying to judge between different sets of models for the present.
The following section summarizes the observational details and image reductions 
for both the WFC3/IR near-IR and ACS/WFC optical data.
Section~\ref{sec:phot} describes our photometric measurements and selection of GC candidates.
Section~\ref{sec:distribs} presents the GC color-magnitude diagrams (CMDs) and color
distributions resulting from our new photometry, as well as mixture modeling
analysis results for the different distributions.
The form of the color-color relation between \IHacs\ and \gIacs\ is discussed in detail
in Section~\ref{sec:curvs}.
In Section~\ref{sec:acsfcs}, we cross-match our photometry against the ACS Fornax Cluster
Survey  (hereafter, ACSFCS; \jordan\ et al.\ 2007) catalogue for NGC\,1399 and discuss
the implications for the underlying metallicity distribution.
The final section places our study in the larger context of GC color and metallicity
studies, before listing our main conclusions.

\medskip
\section{Observational Data Sets}
\label{sec:obs}


NGC\,1399 was observed for one orbit in the F110W and F160W bandpasses of
WFC3/IR in 2009~December as part of \hst\ program GO-11712.  The WFC3 IR Channel
uses a 1024$^2$~pix HgCdTe detector with an active field of view of
$\sim2\farcm05\times2\farcm27$, and a mean pixel scale of about
0\farcs128~pix$^{-1}$ (see Dressel \etal\ 2010 for more information).
A primary goal of GO-11712 is to obtain an empirical calibration for the surface
brightness fluctuations (SBF) method in these two WFC3/IR bandpasses similar to those
derived for ACS in F814W and F850LP (Blakeslee \etal\ 2009, 2010b; Mei \etal\ 2005).
To this end, GO-11712 targeted 16 early-type galaxies in the Fornax and Virgo clusters
over a wide range in luminosity and color.  Another important goal of the project is to
investigate the optical-IR colors of the GC populations in these galaxies, using
primarily the F160W bandpass (which affords the wider baseline) combined with existing
ACS optical data.  NGC\,1399 was one of the first large galaxies observed in this
project, and we have used it to optimize our image processing pipeline.  Analyses of the
SBF properties and GC colors
for the full sample of GO-11712 galaxies will be presented in forthcoming works.

A total of 1197~s of integration was acquired with WFC3/IR in the F160W band.  The data
were retrieved multiple times from the STScI archive as the calibration reference files
for WFC3 were updated.  The photometric results improved markedly after flat fields
produced on-orbit (described by Pirzkal \etal\ 2011) were implemented in the STScI
pipeline.  The results presented here are derived from images retrieved from STScI
in 2011~January.  We combined the individual calibrated WFC3/IR exposures into
a final geometrically corrected image using the Multidrizzle (Koekemoer \etal\ 2002)
task in the
STSDAS package\footnote{STSDAS is a product of the Space Telescope Science Institute,
  which is operated by AURA for NASA.}. 
After experimenting with different Drizzle (Fruchter \& Hook 2002) parameters, we
settled on the square interpolation kernel with a ``pixfrac'' value of 0.8 and a final
pixel scale of 0\farcs1~pix$^{-1}$.  This scale is convenient because it is exactly
twice that of ACS/WFC.

\begin{figure*}
\epsscale{0.8}
\plotone{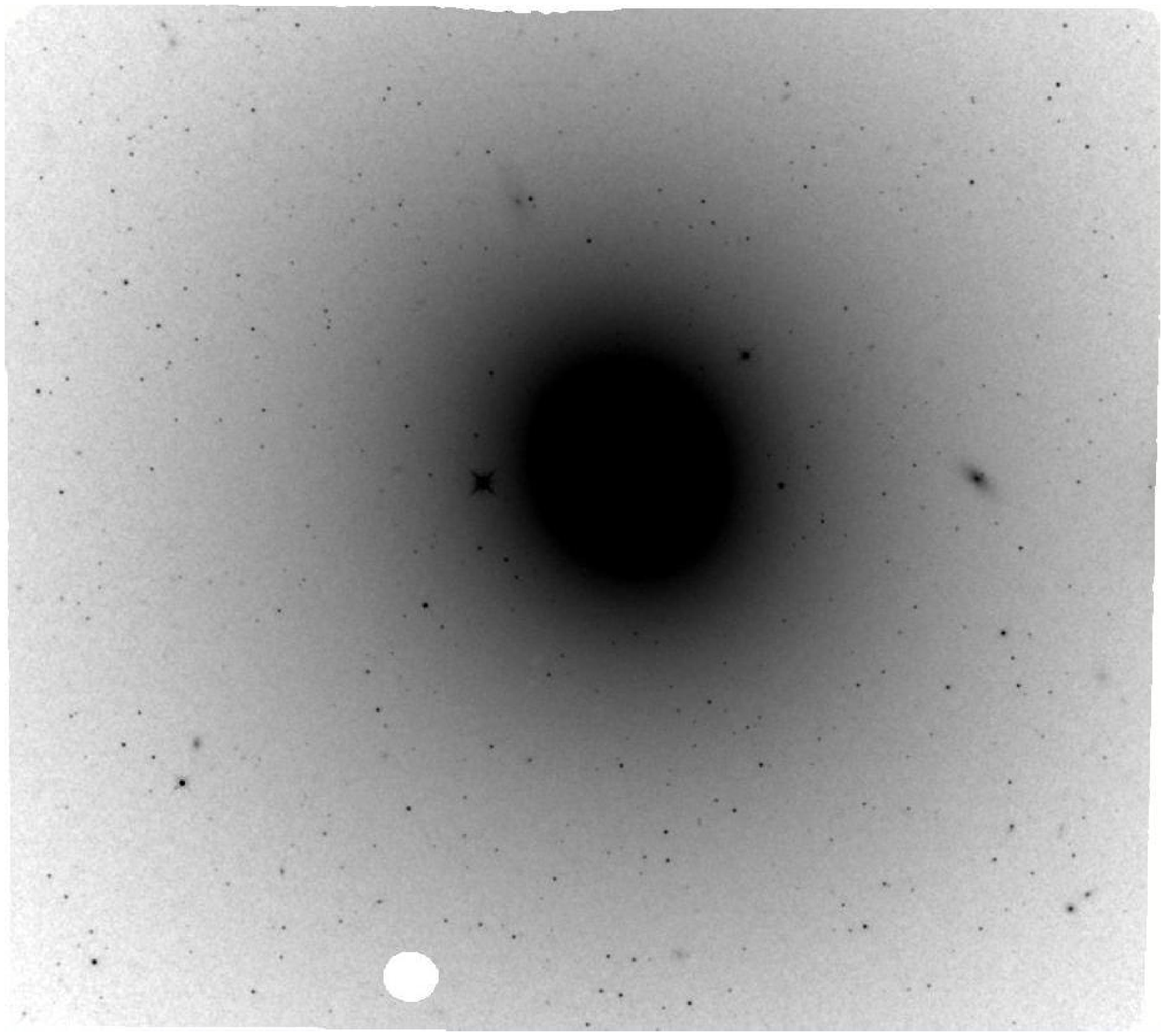}
\plotone{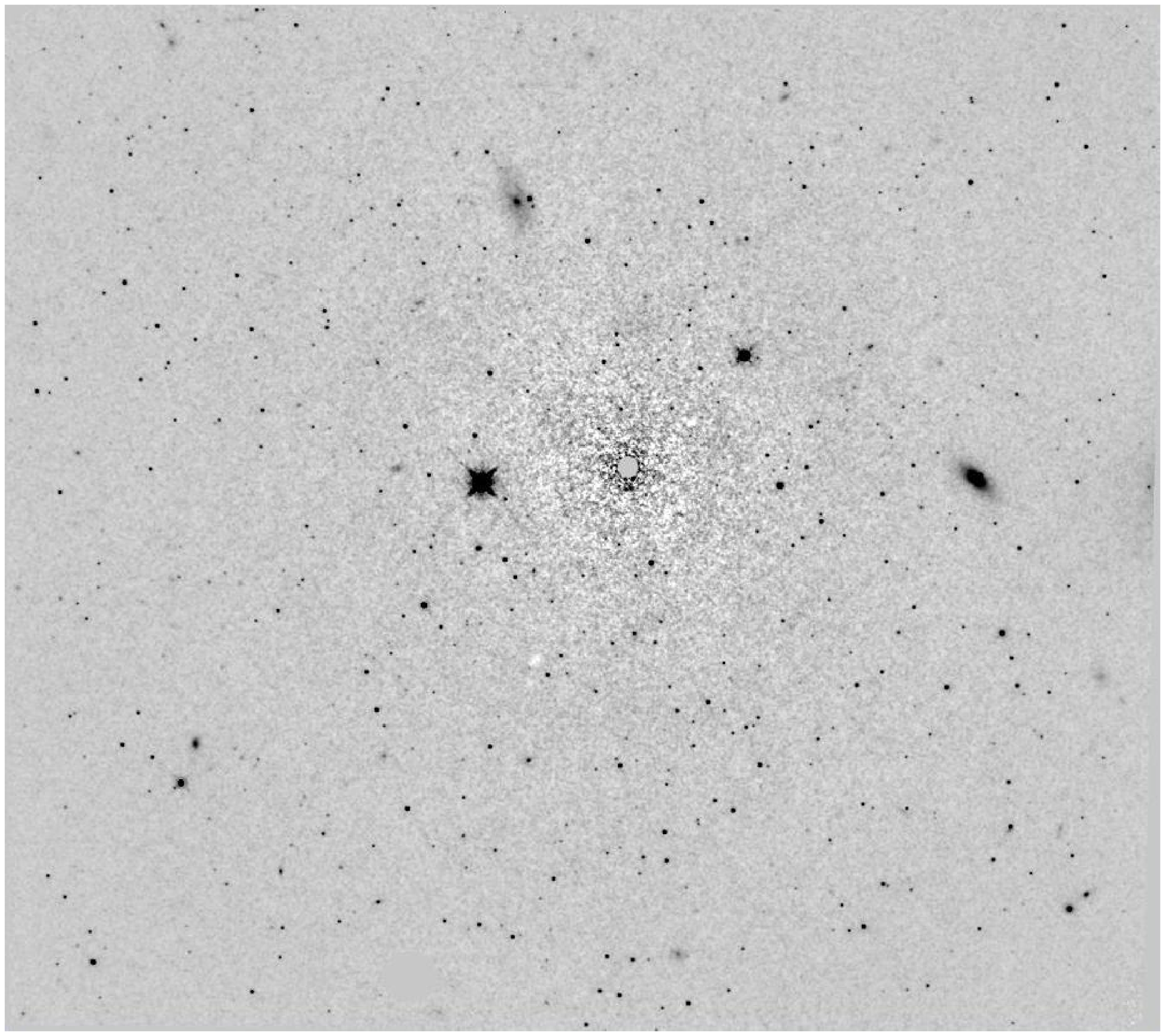}
\vspace{3pt}
\caption{\small
\hst\ WFC3/IR F160W image of NGC\,1399 (top), and the same following
galaxy isophotal model and sky subtraction (bottom).  The field size is approximately
$2\farcm3\times2\farcm1$, and the image is shown at the observed orientation.  
Pixels flagged as bad in the data quality arrays are zeroed in both panels.  
The central 1\farcs2 of the galaxy was not modeled, and is set to zero in the
lower panel.
\label{fig:image}}
\vspace{0.6cm}
\end{figure*}

NGC\,1399 has been observed many times before with \hst, including several times
with the ACS/WFC, allowing us to investigate the optical-IR colors of its GCs.
Calibrated observations in F606W from GO-10129 (PI: Puzia) and F475W$\,+\,$F814W
from GO-10911 (PI: Blakeslee) were retrieved from the STScI archive and processed
with Apsis (Blakeslee et al.\ 2003) to produce summed, geometrically corrected,
cosmic-ray-cleaned images for each bandpass.  Reduction of the GO-10911 imaging
data is described in more detail by Blakeslee \etal\ (2010b); the same procedures
were applied for the GO-10129 program, which observed nine contiguous fields in
F606W. We processed only the four pointings that overlapped with the ACS/WFC
images in other bands; three of these pointings overlap with our smaller WFC3/IR
field.
NGC\,1399 was also observed in F475W and F850LP as part of \hst\ program
GO-10217 (\jordan\ et al.\ 2007); in this case, we used the photometry catalogue
produced by that program, as discussed in Section~\ref{sec:acsfcs}.


Throughout this study, we employ the natural photometric systems defined by the
instrument bandpasses, rather than converting to the Johnson system.
We calibrated the WFC3/IR photometry using the AB zero-point coefficients given
by Kalirai \etal\ (2009) and the ACS photometry using the AB zero~points from
Sirianni et al.\ (2005).  We corrected for Galactic extinction towards NGC\,1399
assuming $E(B{-}V) = 0.0125$~mag (Schlegel et al.\ 1998), the ACS/WFC extinction
ratios from Sirianni et al.\ (2005), and the $H$-band extinction ratio from
Schlegel \etal\ (1998).  
%
Table~\ref{tab:obs} summarizes the observational details of the data sets
used in the present study, including in the last column the symbols used to
denote magnitudes in the various bandpasses.

\medskip
\section{Object Selection and Photometry}
\label{sec:phot}

In order to obtain photometric catalogues of GC candidates, we first constructed
elliptical isophotal models for the different bandpasses as described in our
previous works (e.g., Tonry \etal\ 1997; Jordan \etal\ 2004; Blakeslee \etal\
2009, 2010b) and used these to subtract the galaxy light.  Figure~\ref{fig:image}
shows our WFC3/IR F160W image of NGC\,1399 before and after the galaxy
subtraction.  The compact sources corresponding to GC candidates, as well as
some background galaxies and a few stars, are clearly evident.

\begin{figure}
\epsscale{1.1}
\plotone{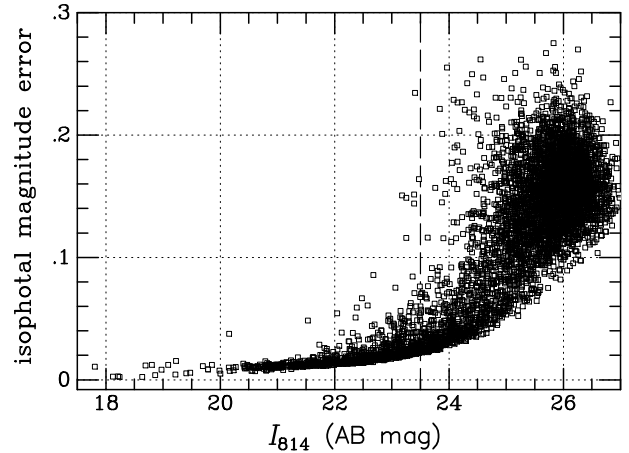}
\caption{\small
The error in magnitude for pixels above the isophotal detection
threshold is plotted against \Iacs\ \magauto\ values from SExtractor.
The magnitudes are on the AB system.
The vertical dashed line at $\Iacs = 23.5$~mag is drawn somewhat fainter than the
mean magnitude
(or ``turnover'') of the GC luminosity function, which occurs at $\Iacs =
23.5$~mag (see text).  The median isophotal
detection error at $\Iacs = 23.5$ is 0.028~mag.
\label{fig:Imag_err}}
\smallskip
\end{figure}

We performed object detection using SExtractor (Bertin \& Arnouts 1996) with an RMS
weight image that included the photon noise and SBF contributions from the
subtracted galaxy.  The full procedure is described in detail by Jordan
\etal\ (2004, 2007) and Barber~DeGraaff \etal\ (2007).  For the most part, we use
F814W to define the sample limits because it is a broad bandpass at the red end of
the optical spectrum, and the signal-to-noise (S/N) of the data is high, despite
the half-orbit integration.  For object detection in the ACS images, we required
an area of at least four connected pixels above a S/N threshold of two; thus, a
minimum S/N of four within the isophotal detection area.  The SBF is quite strong
in the WFC3 F160W image, and we therefore set the detection threshold higher,
requiring a total S/N of at least 6 to avoid spurious detections.  Objects were
detected and measured in each band separately (rather than ``dual image mode''),
then the catalogues were matched across bands; this process removes spurious
objects from the combined multi-band samples.
Figure~\ref{fig:Imag_err} illustrates the depth of the detection in F814W by
plotting the magnitude error within the isophotal area as a function of the
SExtractor \magauto\ parameter, the estimated total magnitude from SExtractor.

Candidate GCs in NGC\,1399 must be nearly point-like
objects in the expected magnitude range.  
Villegas \etal\ (2010) found that the turnover in the GC luminosity function
(GCLF) occurs at $\zacs = 22.802\pm0.044$, where \zacs\ is the total magnitude 
in F850LP from the ACSFCS.  We find in Section~\ref{sec:acsfcs} that the mean color 
of the GC candidates is $\Iacs{-}\zacs \approx 0.3$, where \Iacs\ is the SExtractor
\magauto\ value; this would indicate a turnover in F814W of $\sim23.1$~mag.  This
is consistent with a magnitude histogram of the GC candidates selected below,
although based on the Virgo GCLF and the relative distances of the two clusters,
the GCLF turnover would be expected to occur about 0.2~mag fainter.  In any case,
this is still brighter than the dashed line shown at $\Iacs = 23.5$~mag in
Figure~\ref{fig:Imag_err}.

\begin{figure}
\epsscale{1.1}
\plotone{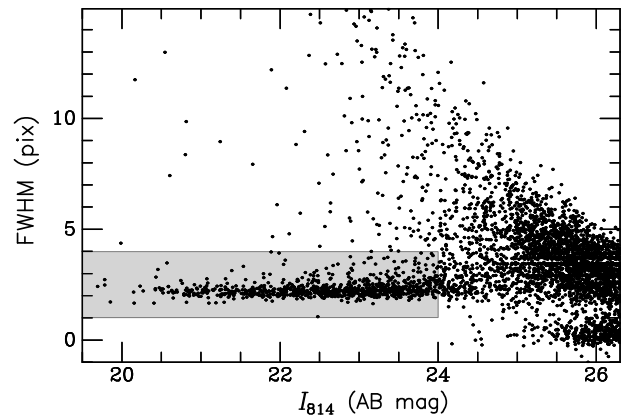}
\caption{\small
Full-width half maximum (FWHM) is plotted against \Iacs\ \magauto\ parameter
from SExtractor for all objects detected in the ACS/WFC F814W image.  The FWHM
values are in pixels, at a scale of 0\farcs05\,pix$^{-1}$.  The gray shaded region
shows the initial selection for GC candidates in this plane.
The turnover of the GCLF occurs near $\Iacs\approx23.2$~mag.
\label{fig:fwhm}}
\medskip
\end{figure}


Figure~\ref{fig:fwhm} shows the full width at half maximum (FWHM) values measured
with SExtractor as a function of \Iacs\ magnitude.  There is a ``finger'' of
compact sources with FWHM $\sim2$ pix that can be distinguished from the
background population down to $\Iacs\approx24$~mag (shaded region).  This is
similar to the FWHM of the point spread function, which is about 0\farcs09, or
1.8 to 1.9~pix.  At the 20~Mpc distance of Fornax, 1\arcsec\ corresponds to about
96~pc, so the typical GC half-light radius of $r_h \approx\,$3~pc corresponds to
0.6~pix for the ACS/WFC.  Thus, the GCs are very marginally resolved, and
some with smaller $r_h$ will be indistinguishable from point
sources.  For our initial GC candidate selection, we therefore selected all
objects in the magnitude range $19.5<\Iacs<23.5$ mag, with $1<\hbox{FWHM}<4$.  We
also require the candidates to be reasonably round, with ellipticity $< 1/3$ (in
practice, this rejects very few objects, since they are already required to be
compact).  In Section~\ref{sec:acsfcs}, where we match our sample against that of
the ACSFCS, we confirm that the overwhelming majority of the objects are
high-probability GCs.

The catalogs produced by SExtractor include magnitudes measured within many
different apertures.  In general, we found that the magnitudes of GC candidates
measured within the SExtractor 6~pix (diameter) aperture proved to be optimal, in
the sense that the scatter between colors was near the minimum, yet the aperture
enclosed $\sim75$\% of the total light for point-like objects, thus making a good
compromise between statistical and systematic errors (cf.\ Appendix~F of Sirianni
\etal\ 2005).  This proved to be the case for both the ACS/WFC and WFC3/IR
photometry.  Although for the ACS/WFC this aperture corresponds to a radius
$r=0\farcs15$, while for the drizzled WFC3/IR data it is $r=0\farcs30$, the 
FWHM of the point spread function (PSF) of WFC3/IR is roughly twice that of ACS/WFC;
thus the aperture corrections 
from Sirianni \etal\ (2005) and Kalirai \etal\ (2009) are very similar when the
WFC3/IR aperture is twice that of the ACS/WFC aperture in angular units.  As
a result, the systematic offsets in the aperture colors due to differential PSF
effects across bands are relatively small.

Following \jordan\ \etal\ (2009), we calculated the color corrections
due to differential
aperture effects within our apertures ($r=0\farcs15$ for ACS/WFC; $r=0\farcs30$ for
WFC3/IR) for a typical GC
(King model with $r_h=3$~pc and concentration $c=1.5$) convolved with the PSF.  For
\gIacs, \VIacs, \VHacs, and \IHacs, the estimated corrections 
are: $+0.02$, $+0.01$, $-0.01$, and $-0.02$ mag, respectively.
Since these offsets are small, systematic, and systematically uncertain 
at the $\sim0.02$ mag level,
we have not applied them to our colors for this analysis, 
but simply note that such offsets would be expected
for some external comparisons.  The conclusions of this work would not
change.  In contrast, the corresponding corrections would be $\sim0.1$~mag for
colors involving \zacs, since the PSF is significantly broader in that band;
however, in that case (see Section~\ref{sec:acsfcs}), we use the ACSFCS photometry,
for which the colors have been converted to the infinite-aperture values assuming
the King model profile for a typical GC convolved with the PSF (see Jordan \etal\ 2009).

\smallskip
\section{Color Distributions}
\label{sec:distribs}

NGC\,1399 was one of the first galaxies suggested to show evidence for
bimodality in its GC metal distribution, based on Washington system
photometry (Ostrov \etal\ 1993), although there was no clear separation
between the proposed components in the color distribution of that early
sample.  The bimodality result, again based on photometric colors, was
more evident for some other galaxies (e.g., Zepf \& Ashman 1993).  It is
now clear that the situation in NGC\,1399 is more complex.  Color bimodality
is present in the system, but not for all subsamples, and we first
address this issue.

\begin{figure}
\epsscale{1.12}
\plottwo{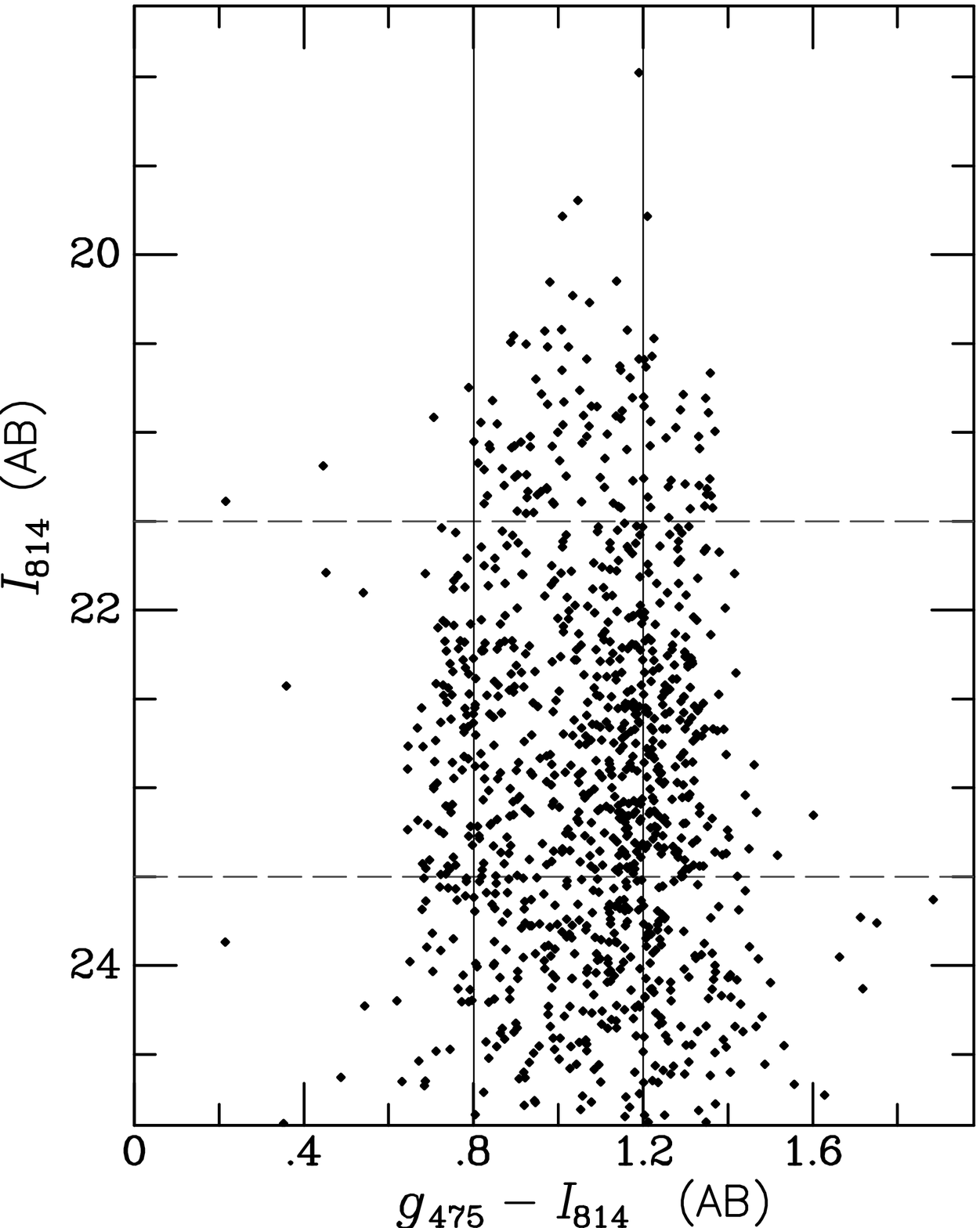}{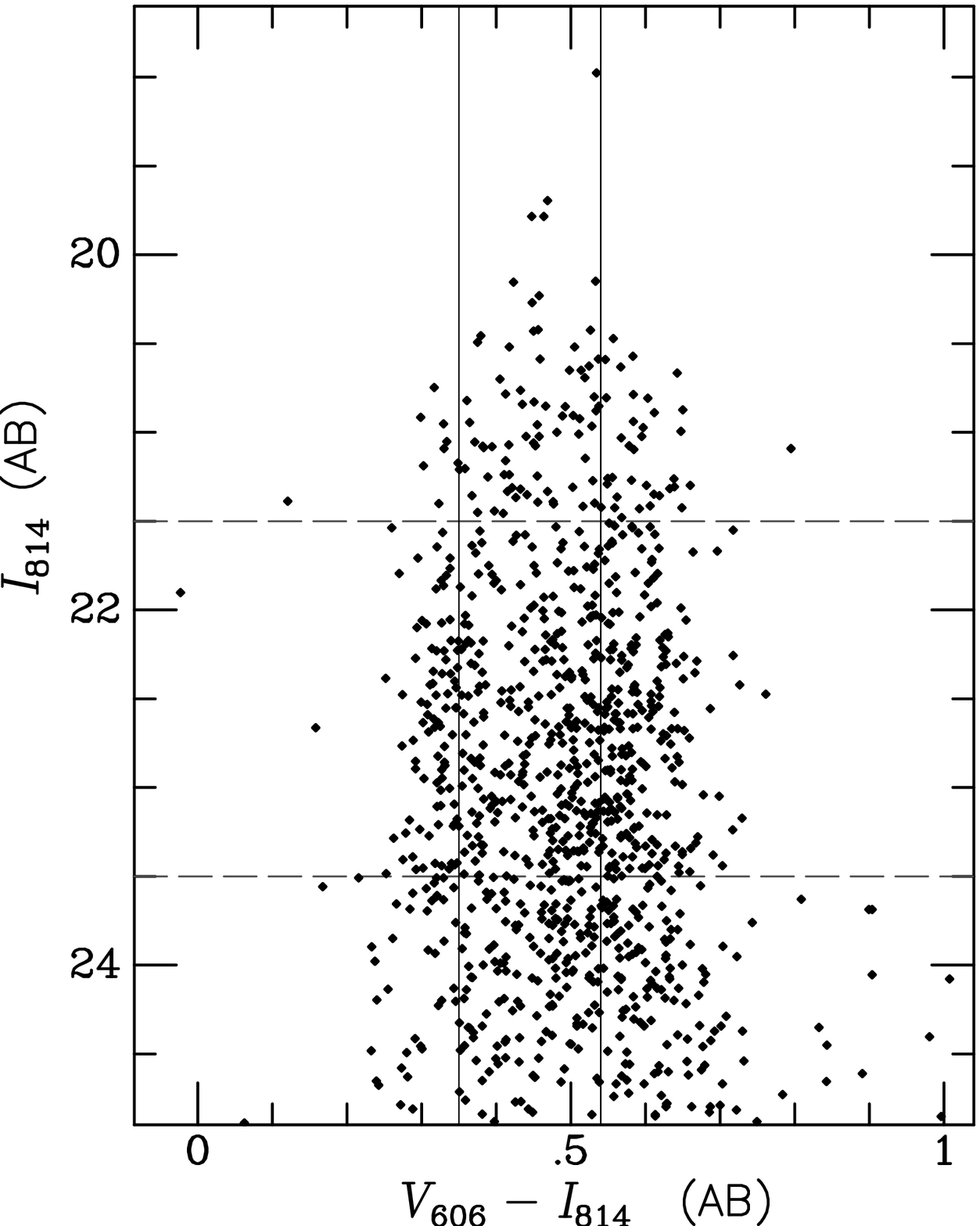}
\vspace{1pt}
\caption{
Optical color-magnitude diagrams for GC candidates in NGC\,1399 from ACS/WFC
imaging.  Dashed horizontal lines indicate the range over which the color
histograms are significantly bimodal, and the vertical lines indicate the peaks in the
color distributions for this magnitude range.  
The brightest GCs, more than $\sim\,$2~mag above the turnover, do not exhibit
distinct bimodality because NGC\,1399 is a strong ``blue tilt'' galaxy.
\label{fig:cmds}}
\vspace{3pt}
\end{figure}

\subsection{Optical Colors by Magnitude}

Dirsch \etal\ (2003) discussed the ``striking'' difference in the color
distribution of NGC\,1399's brightest GCs compared to those within $\lta2$~mag
of the GCLF turnover, based on wide-field CTIO~4\,m Mosaic imaging
extending to $\sim20\arcmin$. The GCs in their brightest magnitude bin followed
a unimodal color distribution peaking at $C{-}T_1\approx1.5$ in the Washington
system, whereas GCs in the next two magnitude bins exhibited a clearly bimodal
color distribution with the ``gap'' occurring near the same $C{-}T_1\approx1.5$
color.  This confirmed the earlier result based on a smaller sample by Ostrov
\etal\ (1998) that the brightest $\sim1$~mag of the GC population had a broad
color distribution, with a mean color similar to the location of the gap in the
distribution of the fainter GCs.  Dirsch \etal\ also found that the red peak was
only prominent at $r\la9\arcmin$, and is more of a red tail at larger radii.
Bassino \etal\ (2006) analyzed two additional Mosaic fields to extend the areal
coverage of NGC\,1399 GC system even further, and found results similar to those
of Dirsch \etal\ (2003).


Figure~\ref{fig:cmds} shows our \gIacs\ and \VIacs\ versus \Iacs\
color-magnitude diagrams (CMDs) for GC candidates selected in \Iacs, as described
above, and matched with objects detected in the GO-10911 \gacs\ data
(taken at the same pointing and within the same orbit as \Iacs) and, separately, with
objects detected in \Vacs\ from the multi-pointing GO-10129 observations.
The colors are measured within the adopted $r=3$~pix aperture, and we only
consider objects with color errors $<0.2$ mag for this aperture.  
Many large galaxies exhibit a ``blue tilt'' in their CMDs, a tendency for GCs in
the blue peak to become redder at higher luminosities, sometimes merging
together with the red component (Harris \etal\ 2006; Mieske \etal\ 2006; Strader
\etal\ 2006; Peng \etal\ 2009). In this sense, NGC\,1399 is a very strong ``blue tilt galaxy,''
since the two color components merge for the brightest GCs (Dirsch \etal\ 2003;
Mieske \etal\ 2010). However, Forte \etal\ (2007) find that for fainter GCs, in
the luminosity range where the distribution is clearly bimodal, there is no
significant slope in the color of the blue peak with magnitude.  This is
consistent with models in which the tilt is due to self-enrichment and
requires some minimum GC mass threshold (Bailin \& Harris 2009).  
Blakeslee \etal\ (2010a) showed that a simple mass-metallicity relation,
combined with a nonlinear color-metallicity transformation, can produce bimodal
color distributions with a blue tilt in GC populations with broad unimodal
metallicity distributions.
In Figure~\ref{fig:cmds}, the horizontal lines enclose the magnitude range 
in NGC\,1399 where the bimodality is most evident, and the vertical lines show the
approximate locations of the peaks, found below.  The tilting of the blue GCs
towards the red at $\Iacs<21.5$ is clear.


\begin{figure*}
\begin{centering}
\includegraphics[angle=270,scale=0.55]{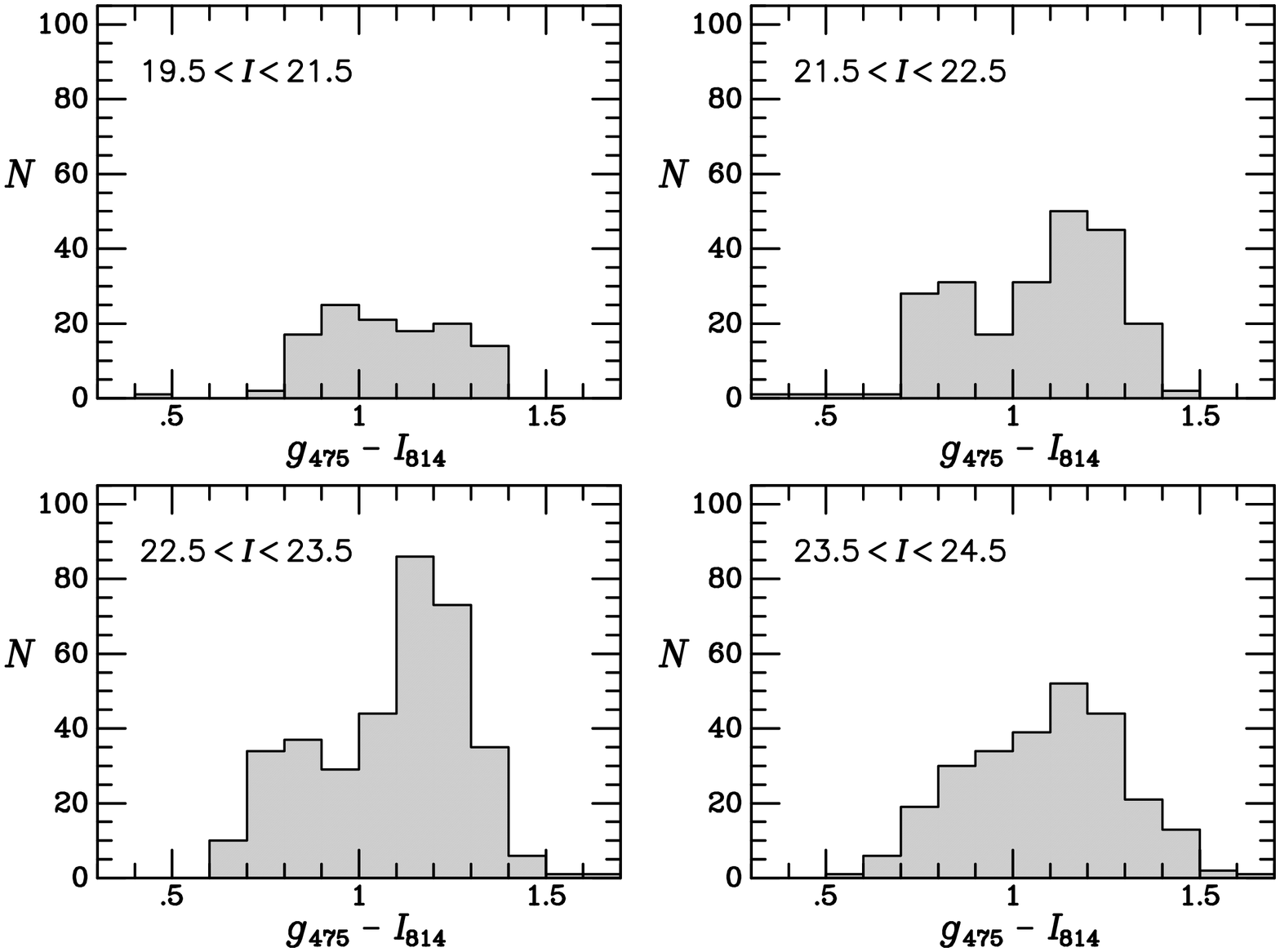}
\medskip
\caption{\small
Histograms of $\giacs$ colors for different \Iacs\ magnitude ranges.
\label{fig:gIbyI}}
\vspace{0.2cm}
\end{centering}
\end{figure*}

\begin{figure*}
\begin{centering}
\includegraphics[angle=270,scale=0.55]{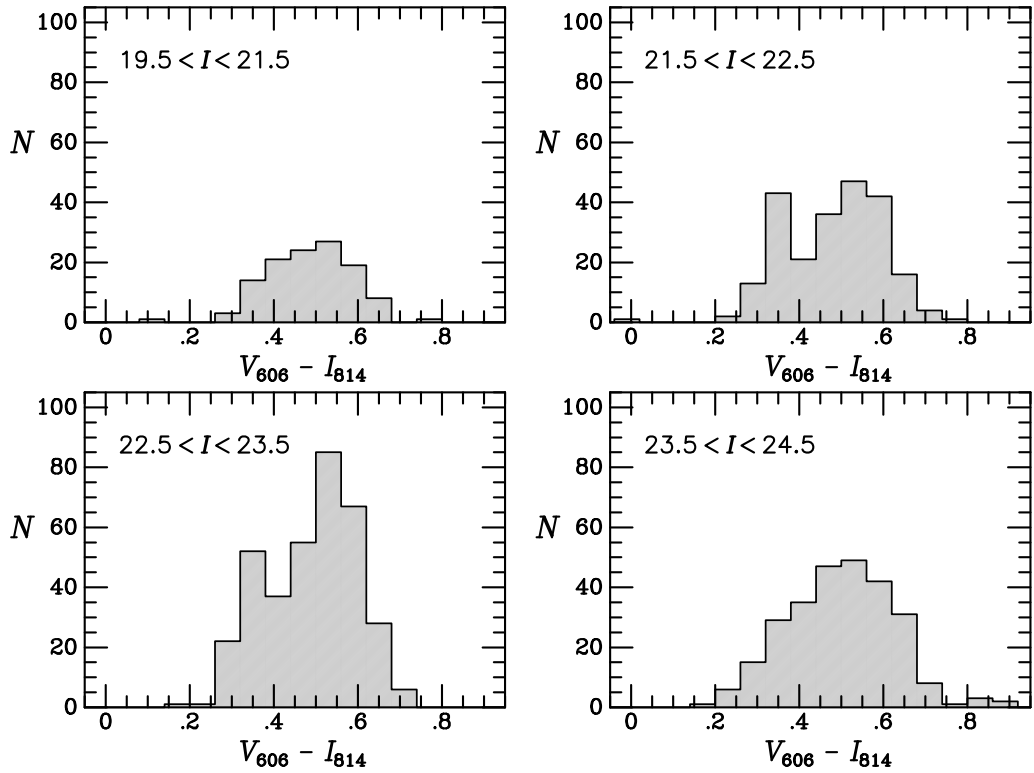}
\medskip
\caption{\small
Histograms of \viacs\ colors for different \Iacs\ magnitude ranges.
\label{fig:VIbyI}}
\end{centering}
\end{figure*}

Figures~\ref{fig:gIbyI} and~\ref{fig:VIbyI} display \gIacs\ and \VIacs\
histograms for GC candidates in several different \Iacs\ magnitude ranges (note
that there are only four brighter than $\Iacs{\,=\,}20$).  In both cases, the
appearance of bimodality is greatest within the $21.5<\Iacs<22.5$ and
$22.5<\Iacs<23.5$ magnitude ranges.  Although the \gIacs\ colors in the
brightest magnitude range of Figure~\ref{fig:gIbyI} appear somewhat bimodal, the
blue ``peak'' in this bright range coincides with the ``gap'' at
$\gIacs\approx0.95$ in the color histograms of the fainter GCs.  The situation
is similar in Figure~\ref{fig:VIbyI}, although in that case, the histogram of
the brightest GCs is essentially a broad red distribution that encompasses the
gap at $\VIacs=0.4$ for the fainter GCs.

Thus, although the Dirsch \etal\ (2003) and Bassino \etal\ (2006) studies
covered a much larger area than our \hst\ data, we find the same lack of obvious
bimodality in the color distribution of the brightest GCs.
Our principal goal here is to examine the optical-IR colors for GCs that define
a distinctly bimodal distribution in purely optical colors.  This provides a
simple, fairly direct test of whether the colors linearly trace metallicity and
reflect true bimodality in the underlying distribution.  In the following
section, we therefore compare the purely optical and optical-IR color
distributions over the $21.5<\Iacs<23.5$ magnitude range where the optical
bimodality is strongest.

\begin{figure}
\epsscale{1.1}
\plottwo{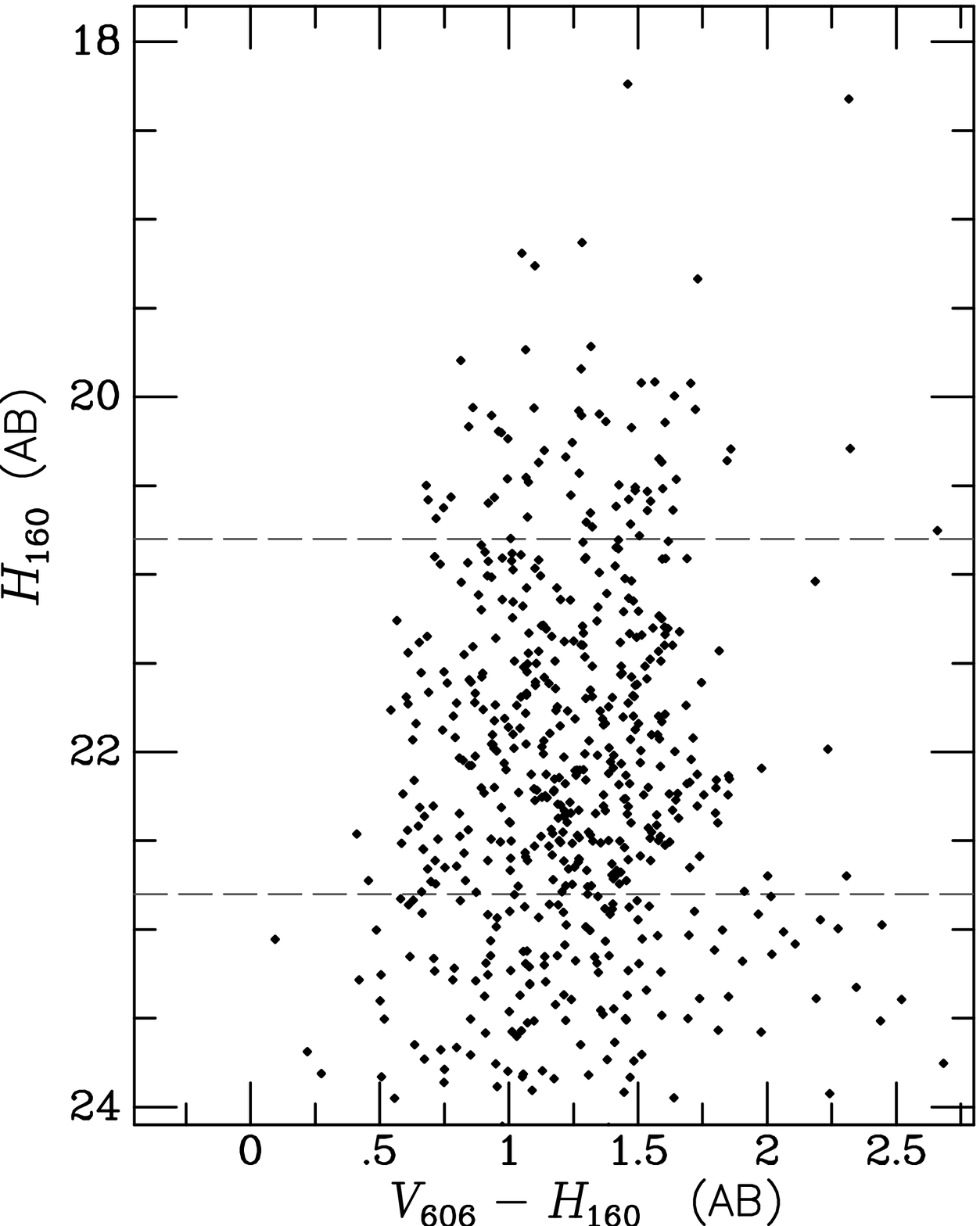}{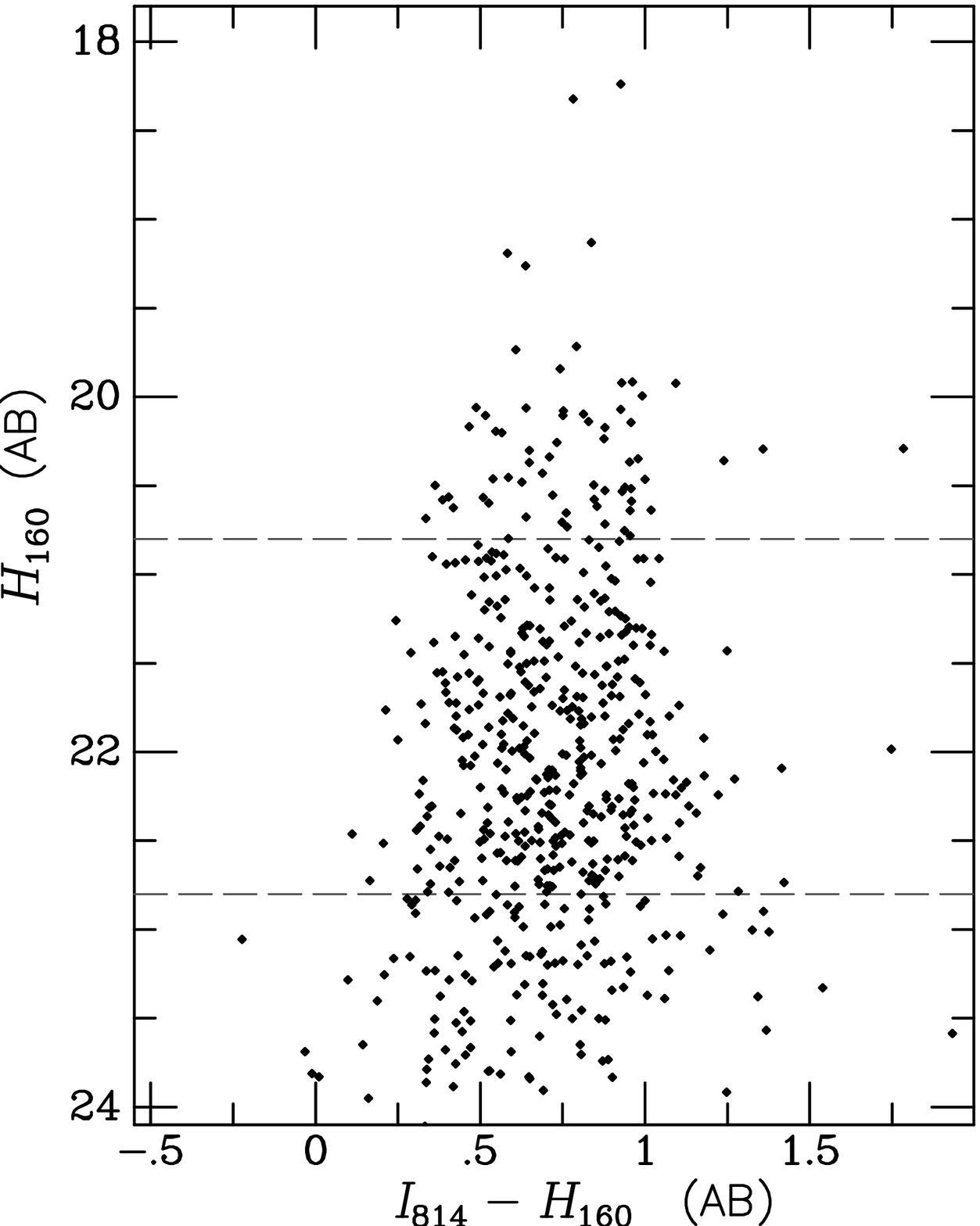}
\vspace{1pt}
\caption{
Optical-IR color-magnitude diagrams.
The dashed horizontal lines are shown to facilitate comparison with the \Iacs\ region
enclosed by similar lines in Figure~\ref{fig:cmds}; they have been shifted by the mean
\IHacs\ color of 0.70~mag.  
\label{fig:Hcmds}}
\smallskip
\end{figure}

\subsection{Optical-IR Color Distributions}


For comparison to Figure~\ref{fig:cmds}, we show the \VHacs\ and \IHacs\ versus
\Hwfc\ CMDs in Figure~\ref{fig:Hcmds}.  The colors are measured within
apertures of $r{\,=\,}3$~pix, as described in Section~\ref{sec:phot}.  Again, only objects
with color errors $<0.2$~mag are shown.  The horizontal lines here are the
equivalent of those shown in Figure~\ref{fig:cmds}, but shifted by the mean
$\langle\IHacs\rangle = 0.70\pm0.01$ mag.  They are merely for comparison to the
previous CMDs, not to select a magnitude range with the most bimodality.  Any
bimodality in these CMDs is much less evident.

\begin{figure*}\epsscale{1}
\begin{centering}
\includegraphics[width=0.45\linewidth]{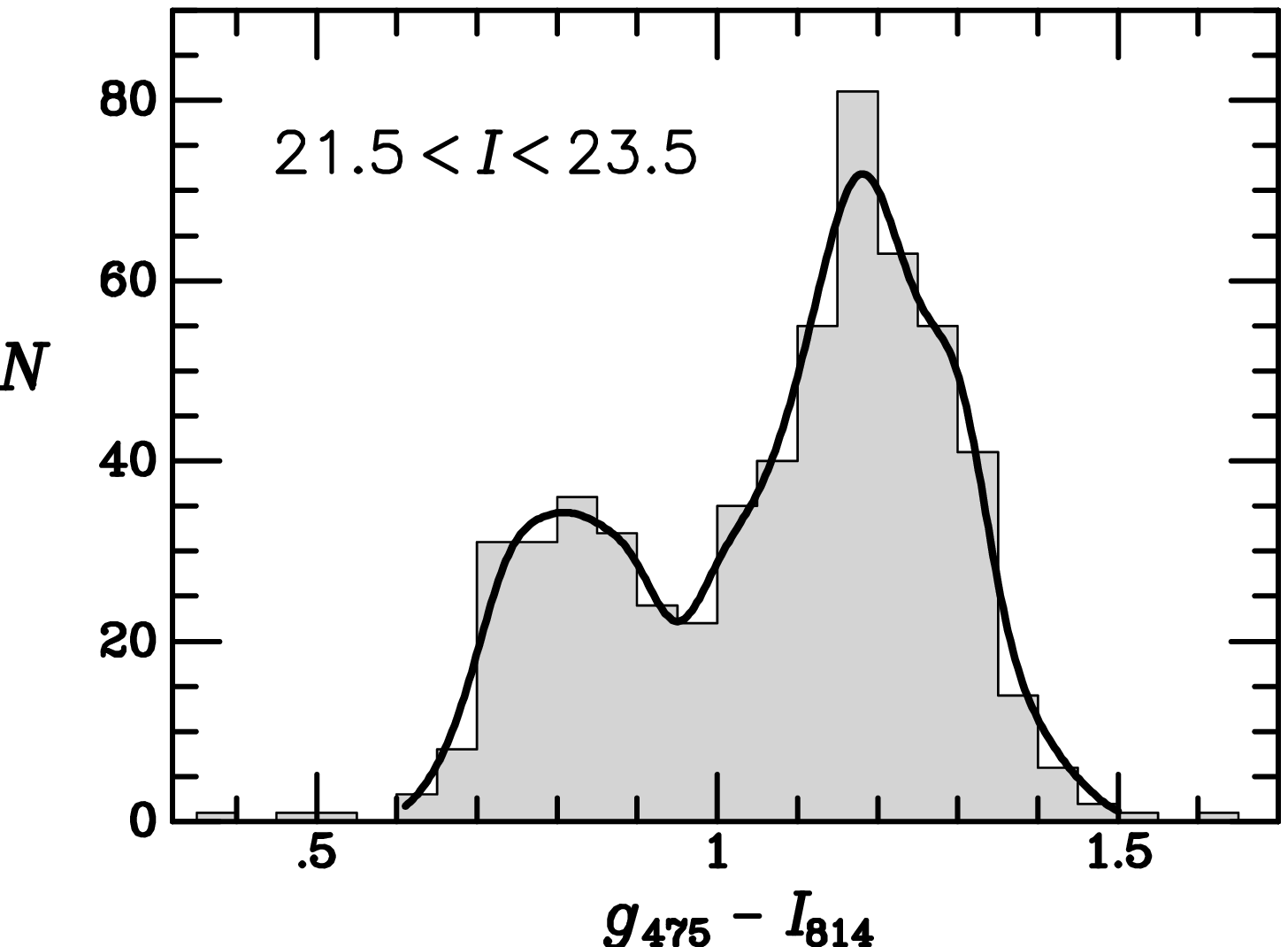}\hspace{0.3cm}
\includegraphics[width=0.45\linewidth]{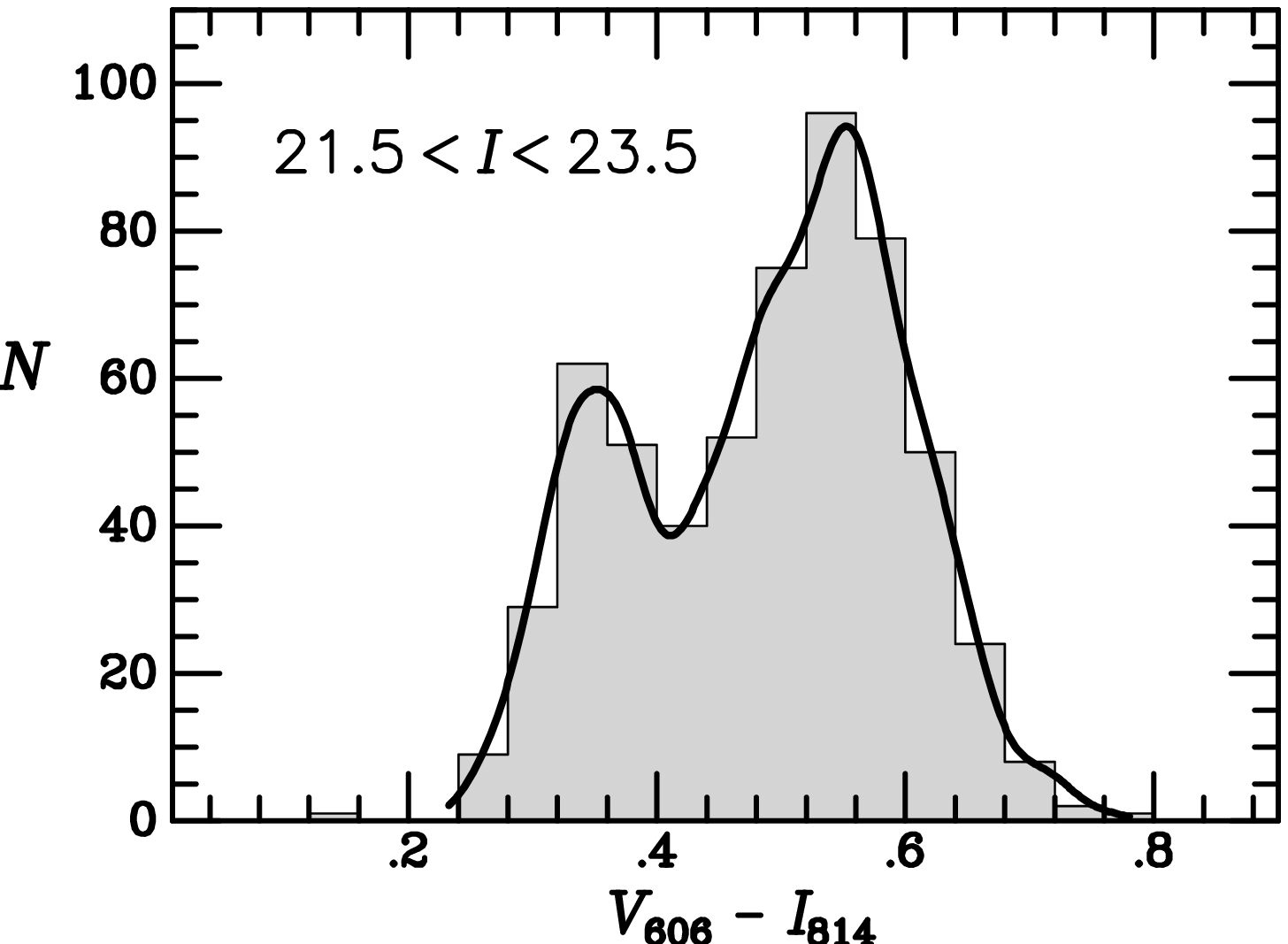} \\[0.5cm]
\includegraphics[width=0.45\linewidth]{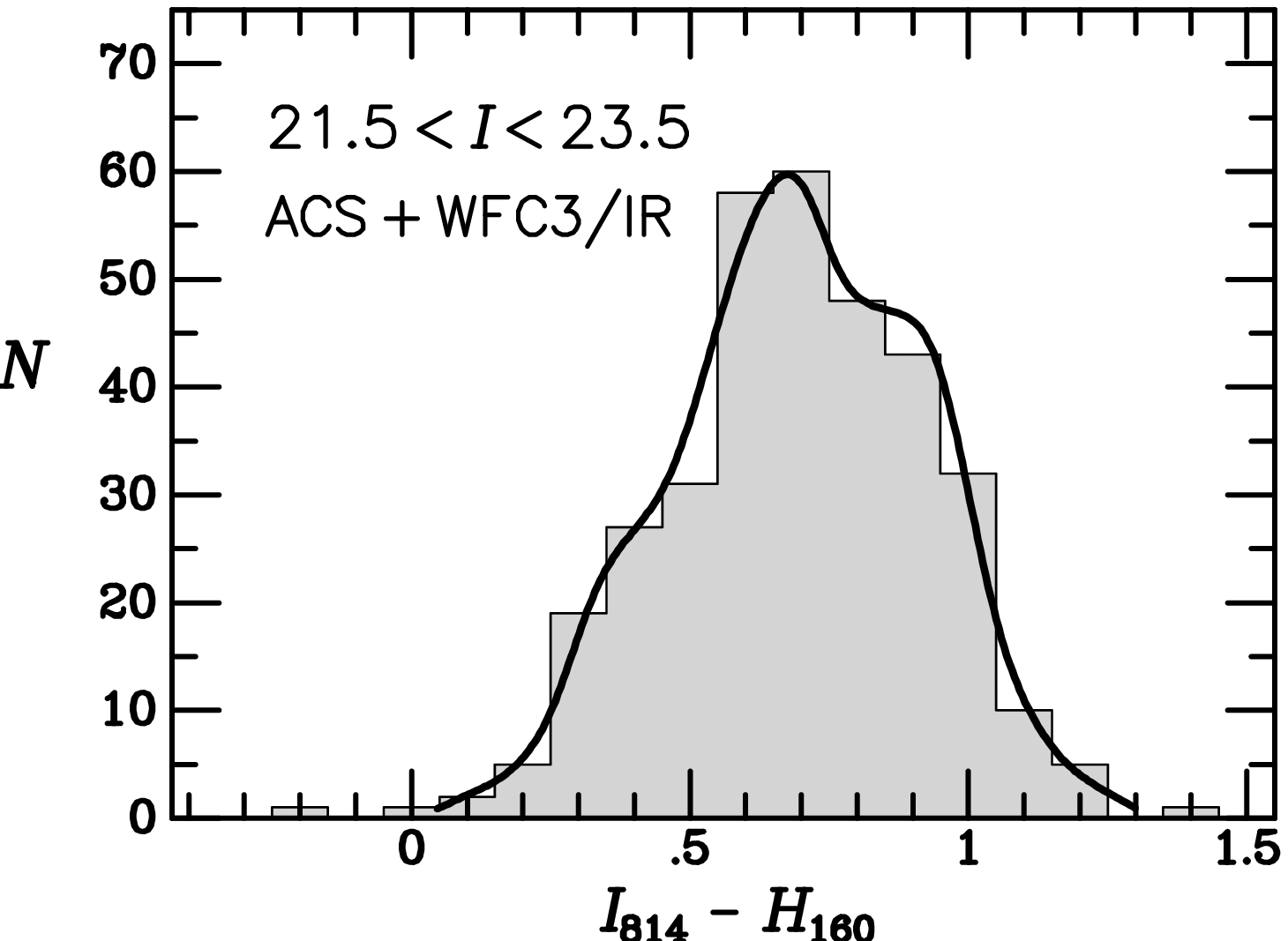}\hspace{0.3cm}
\includegraphics[width=0.45\linewidth]{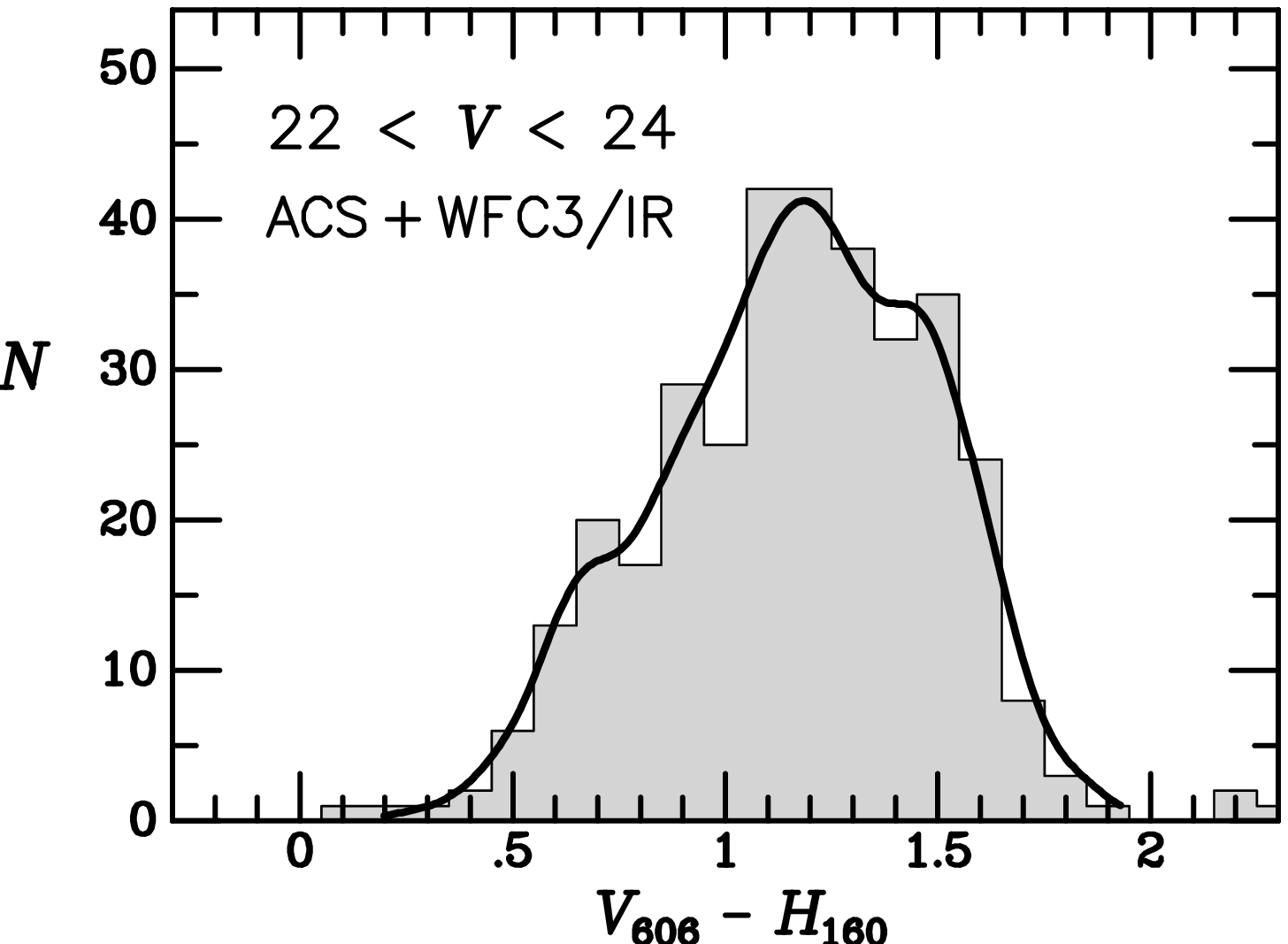} \\[2pt]
\caption{\small
Histograms of \giacs, \viacs, \ihacs, and \vhacs\ colors for all matched GC candidates
in the specified magnitude ranges and with color errors less than 0.2~mag.
\label{fig:4clrhist}}
\end{centering}
\medskip
\end{figure*}

Figure~\ref{fig:4clrhist} shows the binned \gIacs, \VIacs, \IHacs, and \VHacs\
color distributions, along with smooth density estimates constructed with a
Gaussian kernel.  The colors involving \Iacs\ use the magnitude range selected
above, while for \VHacs, we use $22<\Vacs<24$, as the mean \VIacs\ color is
0.49~mag.  We did this for convenience because the pairs of bandpasses were
matched separately at this stage; in Section~\ref{sec:acsfcs}, we examine the
colors of a single merged sample.  In any case, the selection on either \Iacs\
or \Vacs\ makes no difference to the conclusions here.
The \gIacs\ and \VIacs\ color distributions appear strongly bimodal, both having
two distinct peaks with the red peak containing roughly 2/3 of the GC
population.  The predominance of red GCs occurs because the \hst\ fields cover a
relatively small central area, and the NGC\,1399 GC system has a radial color
gradient.
The \IHacs\ and \VHacs\ distributions appear quite similar to each other: unlike
in the optical colors, there are no well-separated peaks, although neither do
they appear to be simple Gaussians.  The \VHacs\ distribution shows some
evidence for a blue component; this is understandable because \IHacs\ is sharply
peaked (though, again, not symmetric), while \VIacs\ is bimodal, so \VHacs\
would be expected to show some bimodality.


We have used the GMM code (``Gaussian Mixture Modeling'') of Muratov \& Gnedin
(2010) to quantify the above qualitative impressions.
These authors provide a detailed discussion of the issues and pitfalls inherent
in bimodality tests.  They note that Gaussian mixture modeling such as GMM or
KMM (Ashman \etal\ 1994), by itself, is more a test of Gaussianity than unimodality. 
As observed by many authors, unimodal but skewed distributions will strongly favor
a double Gaussian model, even if there is no true bimodality in the
distribution.  To address this point, they emphasize the importance of the
kurtosis, stating that ``$\kurt<0$ is a necessary but not sufficient condition
of bimodality.''  This is true because the sum of two populations with
different means is necessarily
broader than a single population; therefore, in the case of Gaussians, which have
$\kurt=0$, the kurtosis must be negative for a valid double Gaussian
decomposition.  In addition, following Ashman \etal\ (1994), Muratov \& Gnedin
define the quantity $D$ as the separation between the means of the component
Gaussians relative to their widths; they state that Gaussian splits that appear
significant based on the $p$ value but have $D<2$ are ``not meaningful,'' in the
sense that they do not demonstrate clear bimodality, but non-Gaussianity.

Table~\ref{tab:gmm_indsamp} presents our GMM analysis results; the errors on the
tabulated quantities come from the bootstrap resampling done in the GMM code.
Note that the analysis is independent of any binning.
The full \gIacs\ sample for the $21.5<\Iacs<23.5$ magnitude range contains 584
objects and strongly prefers a double Gaussian model with peaks at $0.83\pm0.02$
and $1.19\pm0.01$ mag.  (We remind the reader that the small offsets for
aperture corrections, as given in Section~\ref{sec:phot} should be applied for
external comparisons.)  About 70\% of the GC candidates are assigned to the red
component.  The kurtosis is negative, and $D$ is significantly above 2,
confirming a valid bimodal Gaussian model.  We therefore assign a ``Y'' in the
last column of Table~\ref{tab:gmm_indsamp} to indicate the affirmative on the
question of bimodality.  However, because the \gIacs\ distribution has a few
outliers in tails, and mixture modeling codes are generally sensitive to
extended tails, we ran another test after restricting the range to
$0.6<\gIacs<1.5$.  The results, given in the second row of
Table~\ref{tab:gmm_indsamp}, are very similar to prior run, but with the
kurtosis being even more negative and the model uncertainties slightly reduced.
The situation is very similar for \VIacs, which favors a well-separated double
Gaussian model with peaks at  $0.34\pm0.01$ and $0.54\pm0.01$ mag, and 
$74\pm3$\% of the GCs assigned to the red peak.  Again, we answer ``Y'' to the
question of bimodality.

Consistent with the visual impression, the GMM code finds that the \IHacs\
distribution is not bimodal.  When run on the full distribution in
Figure~\ref{fig:4clrhist}, the few objects in the tails result in a positive
kurtosis and virtually all of the objects are assigned to the red peak.  When we
restrict the color range to $0.1<\IHacs<1.3$, the kurtosis becomes negative as
required, but the $p$ value of the double Gaussian model is not significant, and
the fraction of objects assigned to the second component is consistent with zero.

For \VHacs, the situation is a bit more nuanced.  For GC candidates in the broad
color range $0<\VHacs<2$, the kurtosis is negative, and the $p$ value of 0.01 is
marginally significant, but the separation $D$ is not meaningful.  Moreover, the
weakly favored decomposition has only $39\pm22$\% of the GCs in the red peak, as
compared to $\sim70$\% for the optical decomposition.
If we remove the two bluest outliers, the $p$ value becomes slightly less
significant, but now $D$ is marginally above 2, and the fraction of objects in
the second (red) peak is consistent with that found in the optical, although it
is also consistent with 100\% at the 2-$\sigma$ level.  We therefore answer an
uncertain ``Y?'' to the question of bimodality here, consistent with unimodality
in \IHacs\ but bimodality in \VIacs.

As another test of the bimodality in the color distributions, we ran the ``Dip''
test (Hartigan \& Hartigan 1985) supplied with GMM, which nonparametrically
measures the significance of any gaps, or ``dips,'' in a distribution.  Since
the Dip test is insensitive to the assumption of Gaussianity, it is much more
robust, but as discussed by Muratov \& Gnedin, the returned significance levels
are much lower than for the parametric GMM analysis.  Probabilities above
$\sim50$\% may be taken as indicative of likely bimodality.  From this test, we
found that the probability for the most significant gaps in the \gIacs, \VIacs,
\IHacs, and \VHacs\ distributions to be real were 75\%, 97\%, 4\%, and 22\%.

Overall, we confirm that the \gIacs\ and \VIacs\ distributions for GCs in
NGC\,1399 are strongly bimodal, with consistent proportions of red and blue GCs,
when the brightest GCs are excluded. The \IHacs\ distribution is not
significantly bimodal, but there is marginal evidence for bimodality in \VHacs\
consistent with the proportions found for \gIacs\ and \VIacs.  This is
understandable if the optical color bimodality is partly the result of a
changing horizontal branch morphology, since the colors involving bluer
bandpasses would be more affected by the behavior of the hot horizontal branch
stars, whereas colors such as \IHacs\ would not.
Thus, some evidence for bimodality in \VHacs\ would be expected, but weaker than
that found for \VIacs\ (Cantiello \& Blakeslee 2007).

\begin{figure*}
\epsscale{1}
\plottwo{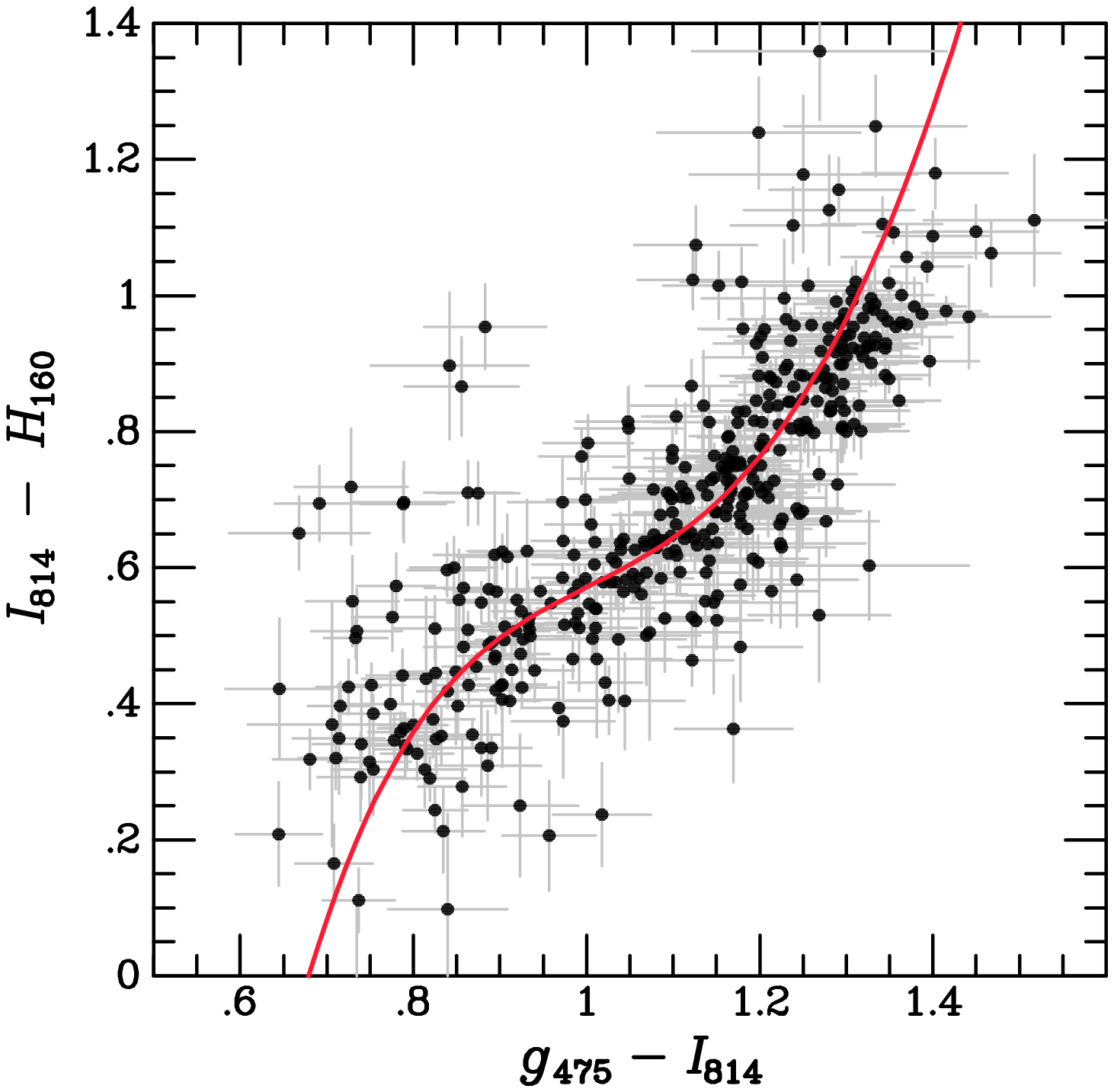}{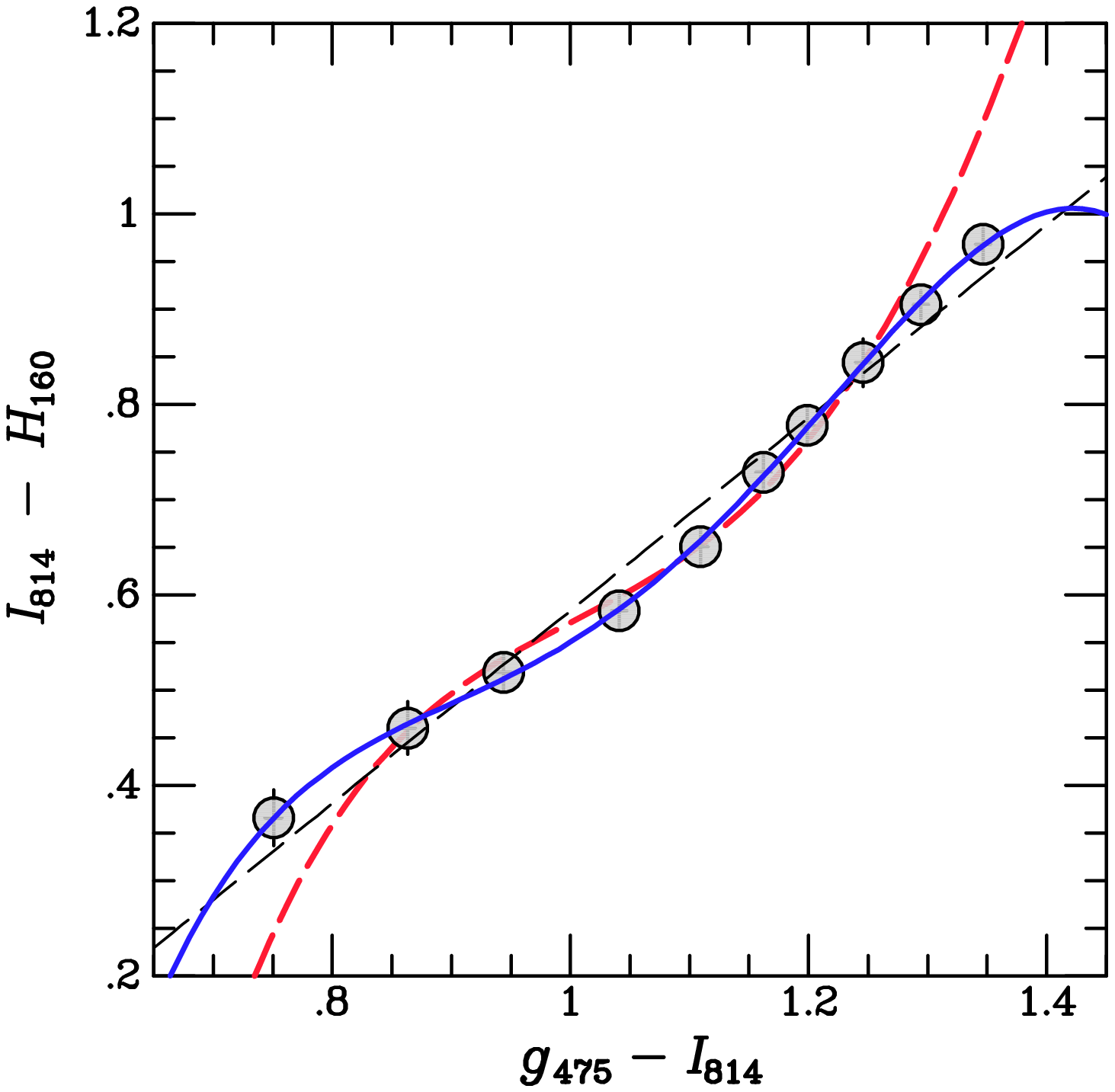}
\vspace{2pt}
\caption{\small
The ACS+WFC3/IR \ihacs\ versus \giacs\ color-color plots.  In the left panel,
the invidual matched GC candidates are shown, with error bars estimated from the
RMS images mentioned in Section~\ref{sec:phot}.  The red curve represents 
a 4th-order robust polynomial fit with coefficients given in the text.  
The right panel shows the median values
within ten bins (40 objects per bin) ordered by \giacs.  We also plot the same
polynomial fit as in the left panel (red dashed curve), a 4th-order polynomial
fit to the median points (blue curve), and a linear fit to these points (black
dashed line).  Error bars on the points are plotted, but in most cases they are
smaller than the point size.  The plot scale is expanded in the right
panel to clarify the differences between the fits.
\label{fig:clr_clr}}
\medskip
\end{figure*}

\smallskip
\section{Color-Color Curvature}
\label{sec:curvs}

The broadest baseline color-color combination we have available for investigating
stellar population issues is \gIacs\ versus \IHacs.  These color indices 
span factors of 1.7--2.0 in wavelength.  The
\gIacs\ index for GCs is affected by the properties of stars near the main sequence
turnoff, on the horizontal branch, and on the giant branch.  It is therefore
sensitive to metallicity, age, and any other parameters (e.g., helium) that
affect the temperature of the main sequence and the morphology of the horizontal
branch.  By comparison, the \VIacs\ index spans a much smaller factor (1.3) in
wavelength, and is less sensitive than \gIacs\ to features 
such as the blue horizontal branch.
As discussed in the Introduction, colors such as \IHacs\ are mainly sensitive to the
temperature of the giant branch, which is controlled by metallicity with very
little age dependence.  Colors such as \gHacs\ and \VHacs\ are also sensitive to giant
branch temperature, but they complicate the metallicity dependence with additional
sensitivity to the main sequence turnoff and horizontal branch.
We therefore concentrate now on the relation between \gIacs\ and \IHacs.


Figure~\ref{fig:clr_clr} (left panel) plots \IHacs\ as a function of \gIacs\ for 401
NGC\,1399 GC candidates in our combined ACS-WFC3 data set with $19.5<\Iacs<23.5$
and $0.5<\gIacs<1.6$.  The corresponding data are listed in Table~\ref{tab:data}.
Although we deal throughout this work with aperture colors within a 3-pixel radius,
in order to facilitate external comparisons, the last two columns of Table~\ref{tab:data} 
provide colors that have been corrected for differential aperture effects,
assuming the corrections for a typical GC, as given in Section~\ref{sec:phot}. 
Since we are interested here in the relationship between
these color indices, rather than simply the presence or absence of bimodality,
we improve the statistics by including the GCs from the brightest magnitude range in
Figure~\ref{fig:gIbyI}, increasing the sample by 17\%.  The relation appears nonlinear.
The plotted curve represents a quartic (4th~order) polynomial, which has been 
fitted to the data using robust orthogonal regression (Jefferys \etal\ 1988).
This approach minimizes the residuals in the direction orthogonal to the
fitted relation, and it is appropriate here
because of the significant observational scatter in both coordinates.
The fitted coefficients are given by
\begin{eqnarray}
\hbox{\IHacs} \;=\; 
-14.54 \,+\, 
 49.23x \,-\, 
 59.02x^2  \,+\, 
 30.14x^3 \nonumber \\
 \,-\; 
  5.24x^4 \,,
\label{eq:empirical}
\end{eqnarray}
where $x\equiv(\gIacs)$.
Several of the objects appear discrepant with respect to this relation: these
may be due to contamination in the sample from Galactic stars and/or background
galaxies, or to real stellar population variations.  In particular, objects that
lie above or to the left of the relation could be young clusters, or they may have extreme
horizontal branches.  Despite some outliers, there is a tight locus of objects
that define the curved sequence in color-color space traced by the above fit.

To further illustrate the curvature, we have binned the data by \gIacs\ into 10
groups of 40 objects, and we plot the median values of the colors within the
bins in the right panel of Figure~\ref{fig:clr_clr}.  The polynomial fit from
the left panel is plotted again, as well as linear and quartic fits to the binned
data.  The deviation from a linear relation is highly significant.  If we use the
scatter to estimate the errors ($\sigma/\sqrt{N}$) in the medians, 
the value of the reduced $\chi^2_\nu$ is 2.7 for the linear fit, and 0.1 for the
quartic fit.  Thus, the linear fit is strongly rejected.
The low value of $\chi^2_\nu$ for the quartic fit probably indicates that the
uncertainties in the medians have been overestimated, most likely because the
small fraction of outliers make the scatter estimate too high.  This also
suggests that $\chi^2_\nu$ should be larger than 2.7 for the linear fit.  
The 4th order model closely traces the curvature of the binned
points; adding another term to the polynomial does not reduce $\chi^2_\nu$
further.  The polynomial fits to the binned and unbinned data diverge near the
endpoints because of the lack of constraints, but they agree well over the 
$\sim\,$0.8--1.3 color range.


\begin{figure}\epsscale{1.14}
\plotone{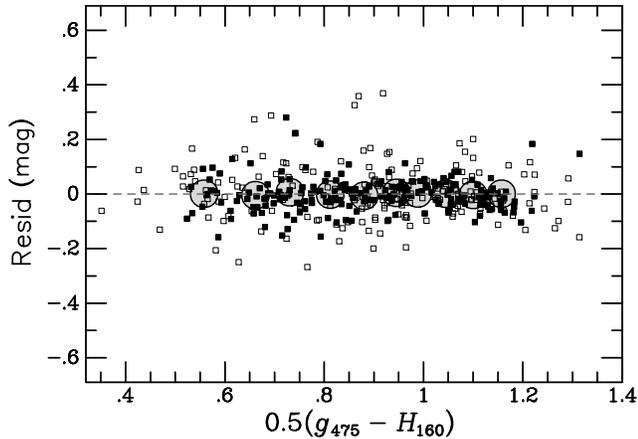}
\vspace{2pt}
\caption{
Color-color fit residuals are plotted as a function of $0.5\times(\gHacs)$, which
corresponds to the mean of \gIacs\ and \IHacs.  We use this mean for the
horizontal axis because it is approximately orthogonal to the residuals that
were minimized in the robust regression fitting (see text).
Square symbols show the orthogonal residuals for individual GC candidates with respect
to the curve in the left panel of Fig.~\ref{fig:clr_clr}.  Filled squares are
used for objects with $\Iacs<22.5$, while open squares are used for objects with
$22.5\le\Iacs<23.5$. 
The large circles indicate the quartic fit residuals for the
median points shown in the right panel of Fig.~\ref{fig:clr_clr}.  
\label{fig:resids}}
\vspace{6pt}
\end{figure}

Figure~\ref{fig:resids} shows the fit residuals as a function of the mean of the
\gIacs\ and \IHacs\ colors, which corresponds to half of the \gHacs\ color.
Since the color-color relation has an average slope of about one
(the linear fit in Figure~\ref{fig:clr_clr} has slope $1.01\pm0.06$), 
this mean color is, to a good approximation, proportional to the distance along
the relation (and therefore in the direction orthogonal to the residuals).
The small squares in this figure represent individual GC candidates, and their 
residuals are with respect to the fit given in equation~(\ref{eq:empirical}). 
Open squares are used for the faintest
magnitude of GC candidates included in the fitting, while solid
squares are used for brighter objects.  The fainter objects scatter more,
but otherwise do not differ systematically from the brighter ones.
The large circles in Figure~\ref{fig:resids} show the residuals for the
median-filtered points with respect to the quartic fit for these points plotted as a
solid curve in the right panel of Figure~\ref{fig:clr_clr}.  In this case, the
median values have very little scatter, and the fit is essentially identical
regardless of whether orthogonal regression or a simple least-squares approach
(minimizing residuals in \IHacs) is used.  From Figure~\ref{fig:resids}, we
conclude that the form of the color-color relation is
well approximated by a quartic polynomial over a range in \gHacs\ from
$\sim\,$1.3 to $\sim\,$2.2 mag, which corresponds approximately to the
$0.8\lesssim\gIacs\lesssim1.3$~mag range quoted above.

We emphasize that not only is the relation between \gIacs\ and \IHacs\
nonlinear, but \textit{the slope attains a local minimum} within the color range
where it is particularly well constrained.  In fact, the inflection point of the
quartic fit to the unbinned data occurs at $\gIacs=1.002$, which corresponds
closely to the dip in the \gIacs\ color distribution in
Figure~\ref{fig:4clrhist}.  If \IHacs\ is a good, relatively simple, indicator
of metallicity, in line with theoretical expectations, then this type of
``wavy'' or inflected relation could produce double-peaked optical color
histograms from metallicity distributions that have a very different form.  The
metallicity distribution does not need to be unimodal or even symmetric;
chemical enrichment scenarios tend to produce asymmetric distributions (see Yoon
\etal\ 2011b for a detailed discussion of this issue).  We address the
metallicities briefly in the following section.

The observed wavy relation between \gIacs\ and \IHacs\ accounts for the marked
difference between the purely optical and optical-IR color distributions found
in Figure~\ref{fig:4clrhist}.  The significance of the curvature demonstrates
that the apparent disparity was not the result of underestimated photometric
errors obscuring bimodality in \IHacs.

\medskip
\section{\hbox{ACSFCS Comparison: Implications for Metallicity}}
\label{sec:acsfcs}

The ACSFCS (\jordan\ \etal\ 2007) is a survey with ACS in the F475W and F850LP
bands of 43 early-type Fornax cluster galaxies.  All objects with colors and
magnitudes consistent with GCs in the program galaxies were fitted with King
(1966) models using the methodology of \jordan\ \etal\ (2005) to derive
half-light radii $r_h$ and total magnitudes.  As described in \jordan\ \etal\
(2009), objects with colors in the
range $0.5<\gzacs<1.9$ and radii $0.75<r_h<10$ pc were assigned probabilities \pgc\
of being GCs based on their \zacs\ magnitudes and half-light radii, as
compared with those found for objects in background fields.  Those with
$\pgc>0.5$ were accepted as likely GCs.  
For the \gzacs\ colors considered in this section, we used the NGC\,1399 photometric
catalogue produced as described in \jordan\ \etal\ (2009) and already used in several
ACSFCS publications (Masters \etal\ 2010; Mieske \etal\ 2010; Villegas \etal\ 2010;
Liu \etal\ 2011).

We used the object positions to match the ACSFCS photometry for NGC\,1399
with our \gacs, \Iacs, and \Hwfc\ measurements.
We keep high probability GCs with $\pgc\ge0.9$ (in practice, 
there was only one matched object with $0.5<\pgc<0.9$).
This merged data set provides two important benefits.  First, the
probabilistic selection based on $r_h$ and \zacs\ should remove most of the
remaining contaminants in our sample.  Second, the ACS \gzacs\ colors have been
calibrated empirically against metallicity. Although the throughput for objects
with GC spectra is about twice as high in F814W as in F850LP, allowing more
precise color measurements with \Iacs\ for a given exposure time, the longer baseline
afforded by \zacs\ improves the metallicity sensitivity (\cote\ \etal\ 2004).  Peng \etal\
(2006) found that the relation between \feh\ and \gzacs\ was not adequately
described by a linear fit; they used a broken linear model with a shallower slope
for $\gzacs > 1.05$.  This was adequate over the range $\gzacs = 0.7$--1.4, but
under-predicted the spectroscopic metallicities of M87 and M49 GCs in the 1.4--1.6
color range.  Additional curvature was needed to match these.  To accommodate
the measured metallicities of these redder GCs, Blakeslee \etal\ (2010a) used a
polynomial fit that followed the data over the full 0.7--1.6 range in \gzacs.
This empirical calibration provides us with some handle on the metallicities for our
sample.

\begin{figure*}
\begin{centering}
\includegraphics[angle=270,scale=0.7]{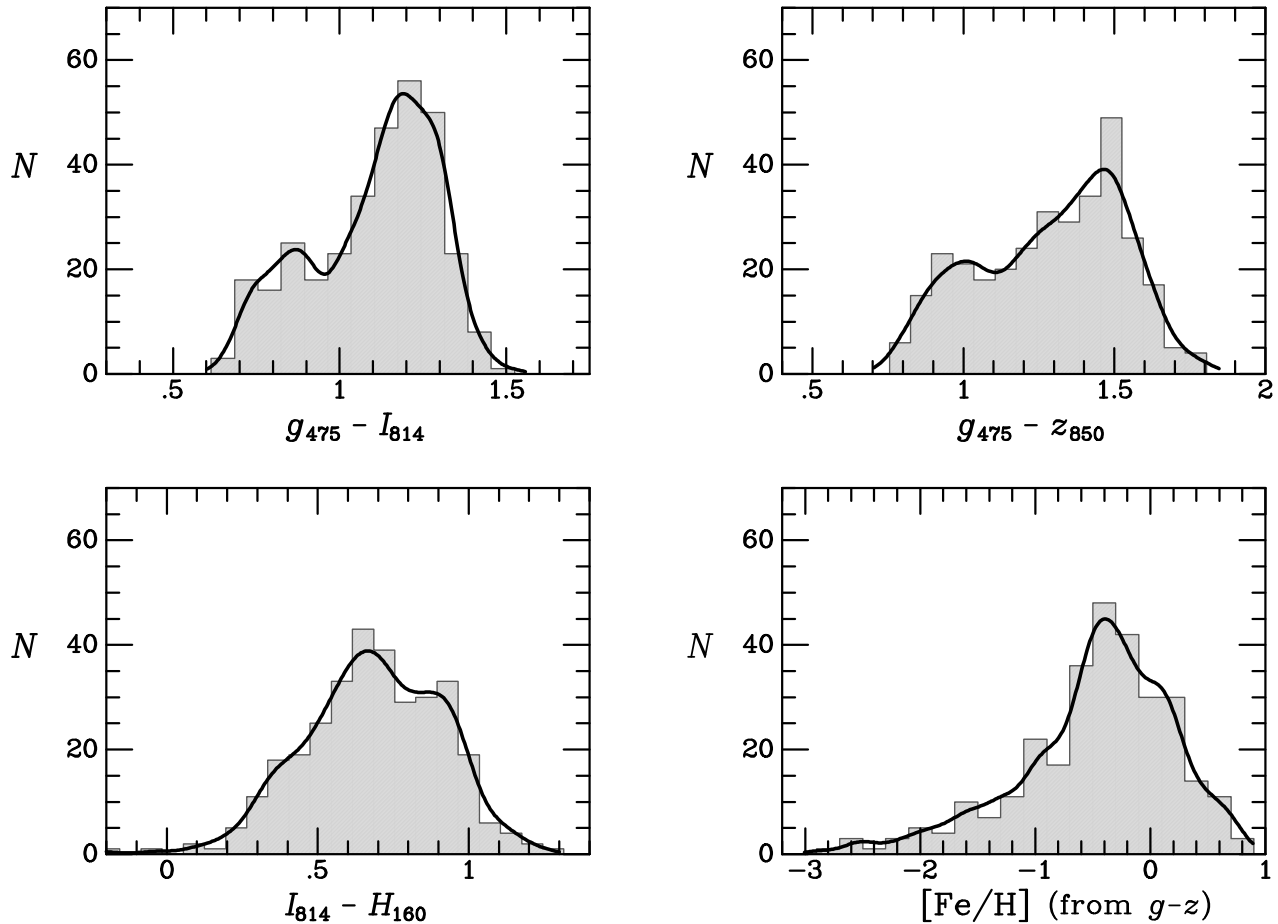}
\vspace{2pt}
\caption{\small
Histograms of \giacs, \gzacs, and \ihacs\ colors, along with \feh\
values  inferred from ACSFCS \gzacs\ data (see text), are plotted for GC candidates
identified in the ACS Fornax Cluster Survey.  
\label{fig:mergehist}}
\end{centering}
\medskip
\end{figure*}


Figure~\ref{fig:mergehist} shows the color and estimated metallicity histograms
for our sample of ACS+WFC3 data after merging with the ACSFCS dataset.  Again we
use the $21.5<\Iacs<23.5$ magnitude range for the histograms to select the
regime of optical color bimodality.
After the merging with ACSFCS, there was one object at $\gIacs=0.45$ mag, whereas
all the others were in the 0.64--1.52 range; we removed this single outlier in
\gIacs\ color.
In contrast to the approach in Figure~\ref{fig:4clrhist}, the exact same 322
high-probability GCs are represented in all of the color histograms in
Figure~\ref{fig:mergehist}.  For uniformity, we have used the same binsize
of 0.07~mag for all the colors.  The change in binning is mainly responsible for
the different appearance of the \IHacs\ histograms in Figures~\ref{fig:4clrhist}
and~\ref{fig:mergehist}, but the kernel density curves reflect the same
features.

As described above, the metallicity histogram in the lower right of
Figure~\ref{fig:mergehist} is based on the polynomial fit by Blakeslee \etal\
(2010a) to the Peng \etal\ (2006) data.  We use this polynomial transformation
only for objects with $\gzacs<1.6$ mag, since it is not constrained beyond this;
objects with redder \gzacs\ colors (7.5\% of our merged sample) are likely
affected by observational scatter.  Thus, there are fewer objects in the \feh\
histogram, and the truncation at $\feh = 0.76$ dex is artificial.  However, the
peak at $\feh\approx-0.3$ dex, and the long tail to lower metallicities, accurately
reflect the \gzacs\ distribution combined with the empirical metallicity
calibration.

Table~\ref{tab:gmm_mrgsamp} summarizes the GMM results for this merged sample.
For the \gIacs\ distribution, we again find a significant preference for a
double Gaussian model, with peaks at $0.84\pm0.02$ and $1.20\pm0.01$ mag and
$73\pm5\%$ of the objects in the red component.  The result is similar for
\gzacs, with peaks at $0.98\pm0.04$ and $1.41\pm0.01$ mag and
$72\pm7\%$ in the red component.  When the analysis is run on the full
\IHacs\ sample, we find $\kurt > 0$, $D<2$, and $p\gta0.05$; thus, no
significant evidence for a bimodal model.  However, no outliers have been
trimmed from the \IHacs\ distribution; if we remove the two objects with
$\IHacs<0$, then the fourth row of Table~\ref{tab:gmm_mrgsamp} shows
$\kurt<0$, as required for significant Gaussian bimodality, but the $p$ value
increases to 0.055, and $D$ is unchanged, except in the uncertainty.  
If we trim the outliers more aggressively, restricting the range to
$0.2<\IHacs<1.2$, then \kurt\ decreases further, and the separation between the
best-fit peaks is now a marginally significant $D=2.0\pm0.3$, but $p$ increases to
0.062 (significance less than 2$\sigma$).  Moreover, the preferred breakdown has
$24\pm28\%$ of the objects in the red component (before the trimming, the red
percentage was even lower).  This red fraction is consistent with zero; however,
if it were interpreted as real, then this would indicate a ``different
bimodality'' than that found for the optical colors, where $\sim\,$73\% of the
objects are in the red peak.
Thus, the preferred optical and optical-IR bimodalities could \textit{not}
both reflect a common, more fundamental bimodality in metallicity.

\begin{figure*}
\epsscale{0.95}
\plottwo{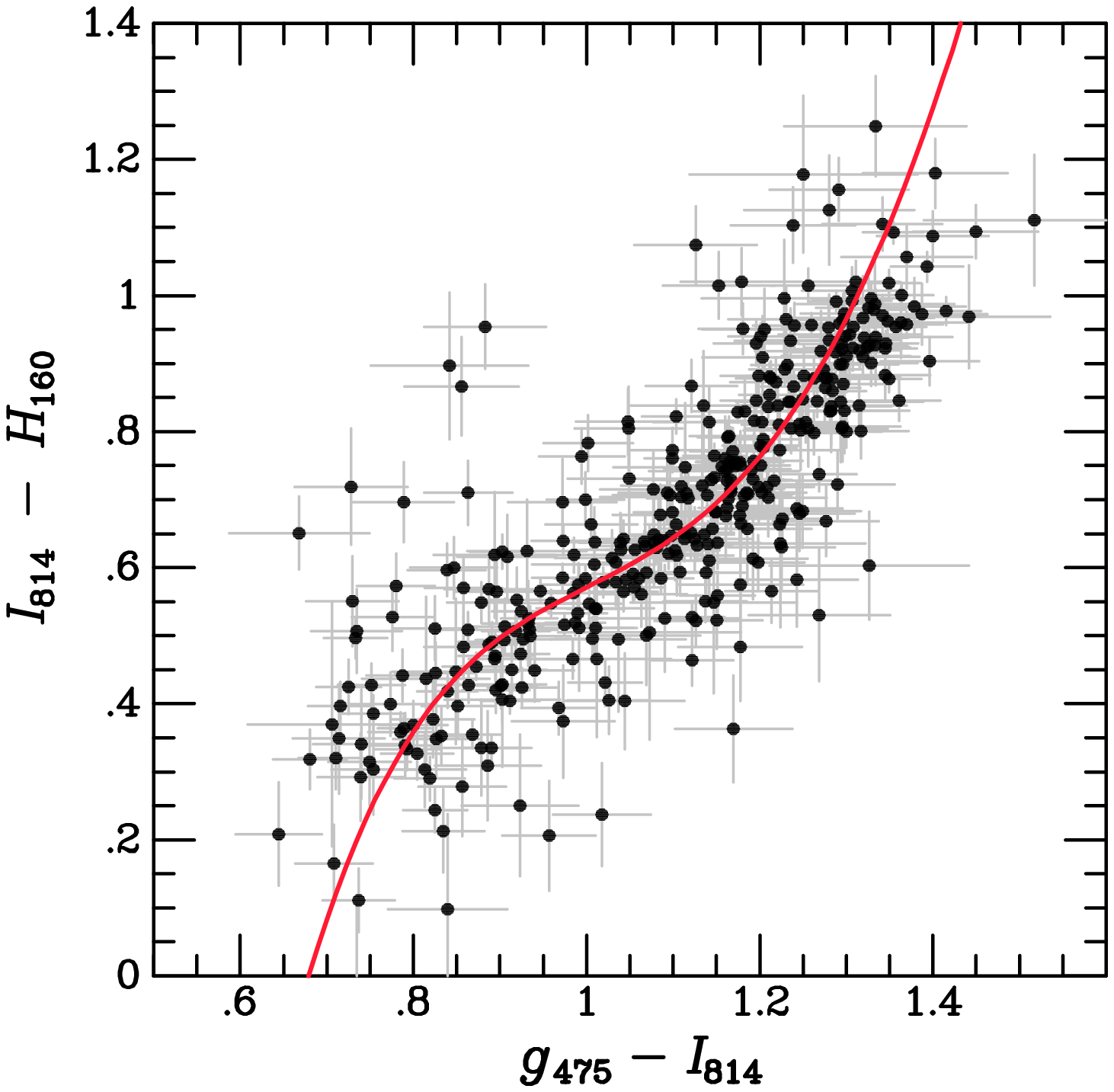}{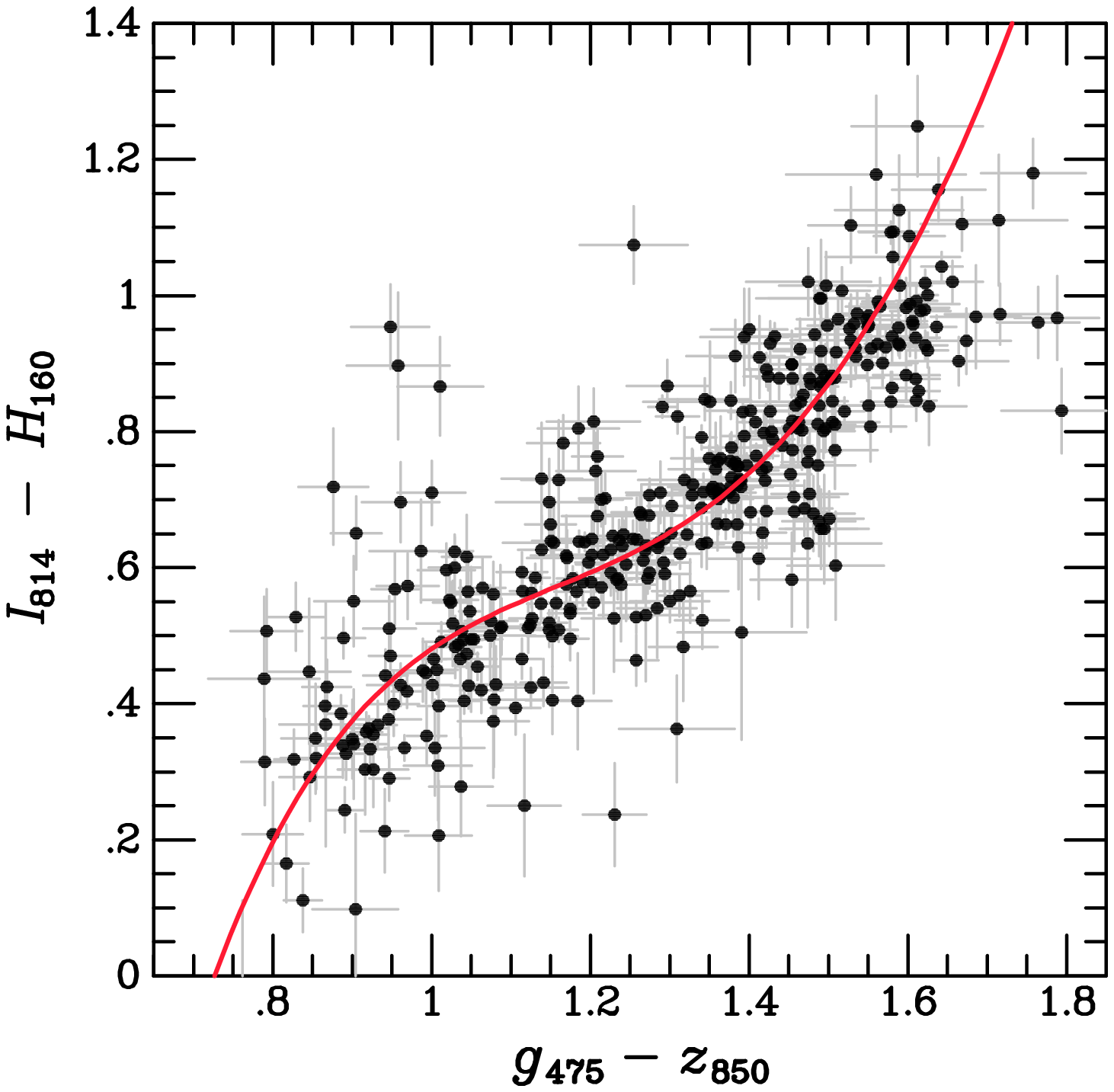}
\vspace{3pt}
\caption{\small
Similar to the left panel of Figure~\ref{fig:clr_clr}, but here shown for the
merged sample of high-probability GC candidates identified in the ACSFCS.
The \gzacs\ measurements are from the ACSFCS.
\label{fig:gI_gz_IH}}
\bigskip
\end{figure*}

We also experimented with the GMM analysis for the \feh\ distribution derived from
the \gzacs\ colors.  For the full sample, the kurtosis is significantly above
zero, and the code wants to fit a primary peak with 80\% of the objects, then use
another broad Gaussian for the blue tail.  This reflects the obvious lack of
Gaussianity.  If we aggressively trim the blue tail, removing objects with
$\feh<-1.9$, then we find $\kurt<0$, $p=0.002$, and $D>2$.  However, the fraction
of objects in the main peak is then $0.917\pm0.084$, consistent with 100\%.  More
to the point, this procedure of removing members of a continuous extended
distribution (rather than outliers), is not statistically valid; we have done so
only for illustration.  The metallicities for NGC\,1399 GCs derived from the
(bimodal) \gzacs\ colors and empirical calibration define a continuous
distribution with a peak at $\sim\,-$0.3 dex and an extended blue tail.  The form of
the high-metallicity end is unknown, but it is unlikely to extend much beyond
the limit imposed here by the calibration.


Figure~\ref{fig:gI_gz_IH} plots \IHacs\ versus \gIacs\ and \gzacs\ for the merged
sample.  This includes 385 objects with $19.5<\Iacs<23.5$; thus, the merging with
ACSFCS removed only 16 objects, or 4\% of the sample.  The resulting plot
for \gIacs\ is essentially identical to that in Figure~\ref{fig:clr_clr}.  The
robust orthogonal regression insured that the rejected objects (possible
contaminants) did not strongly affect the fit.  To illustrate the consistency
between the two relations in Figure~\ref{fig:gI_gz_IH}, we have not independently fitted the
relation for the ACSFCS \gzacs\ colors, but simply transformed the fit for \gIacs\
aperture colors according to the linear relation we find for the data:
\begin{equation}
(\gIacs)_{\rm r{=}3p} \;=\; 0.13 \,+\, 0.75(\gzacs)_{\rm ACSFCS} \,.
\end{equation}
It is not surprising that the relations between \IHacs\ and these two
optical colors would be very similar, since they have the same blue bandpass, and there is
substantial wavelength overlap of the red bandpasses.  We note however that the
\gacs\ measurements are from independent data sets, one taken at the same time and
orientation as the \zacs\ data in \hst\ Cycle~13, and the other taken with the
\Iacs\ data in Cycle~15.


The merged sample analyzed here is nearly 5~times larger than
previous optical-IR color studies of GCs with \hst, and has significantly better
precision than previous ground-based studies.  Clearly, even larger,
high-quality optical-IR samples are needed to explore higher-order features of
the color-color relations presented here.  Deep spectroscopy of representative
portions of the photometric samples is also essential for more detailed
constraints on the underlying metallicity distributions.
We are currently analyzing the full sample of 16 galaxies from our WFC3/IR
program, all of which have existing ACS \gzacs\ data.  The results from that
larger sample will allow us to test the conclusions from the present study.

\medskip
\section{Discussion and Conclusions}
\label{sec:sum}


In this study, we have combined ACS and WFC3/IR photometry to examine optical and
optical-IR color distributions and color-color relations for $\sim\,$400 GCs in NGC\,1399,
the central giant elliptical in Fornax.  Consistent with previous wide-field studies
(e.g., Dirsch \etal\ 2003), we find that the brightest GCs (about 2-3 mag brighter than
the GCLF turnover) in NGC\,1399 have a broad red distribution in optical colors.  Fainter
than this, the optical colors become strongly bimodal, with $\sim70\%$ of the GCs in the
red component.  Any bimodality in the optical-IR colors is much less clear; for \IHacs,
the preferred relative proportion of red and blue GCs also differs from in the optical.

In addition, the relation between pure optical and optical-IR color is significantly
nonlinear.  The slope of the dependence of \IHacs\ on \gIacs\ has a local minimum near
$\gIacs\approx1$ (the polynomial fit has a true inflection point here), corresponding
approximately to the dip in the \gIacs\ distribution; the result is similar for \gzacs.
Since GC color is mainly determined by metallicity, at least one of the color-metallicity
relations must also be significantly nonlinear.  Such nonlinearities can produce bimodal
color distributions even when the metallicity distribution lacks any clear bimodality
(Richtler 2006; Yoon \etal\ 2006; Cantiello \& Blakeslee 2007).  The main requirement is
that the slope in metallicity as a function of color must have a local minimum near the
mid-range of the color distribution (for additional discussion, see Blakeslee
\etal\ 2010a).  Such ``wavy'' nonlinearity for optical colors is predicted by some stellar
population models (Lee \etal\ 2002; Yoon \etal\ 2006, 2011a,b; Cantiello \& Blakeslee 2007),
largely as a result of the behavior of the horizontal branch (still imperfectly modeled).

Further evidence for nonlinearity comes from the observed relation of \gzacs\ colors of
GCs with respect to spectroscopically estimated metallicities (Peng \etal\ 2006).  We
have used a smooth fit to this relation, along with \gzacs\ colors from
the ACSFCS, to estimate metallicities for the GCs in our sample.  Matching with the
ACSFCS, which used an independent, probabilistic method for selecting GCs, also confirms
that the level of contamination in our sample is quite small.  The metallicities derived
from the \gzacs\ colors follow a continuous distribution that peaks at $\sim -0.3$~dex and
extends in a low-metallicity tail to $\feh\approx-3$.  Because of the limited
range of the calibration, the form of the high-metallicity end remains uncertain, but it
cannot have the same long extent in logarithmic space as the low-metallicity tail.  This
skewed distribution, based on the empirical calibration, closely resembles the GC metallicity
distributions derived by Yoon \etal\ (2011b) using their model color-metallicity
relations.  As these authors have pointed out, the resulting distributions are in
much better agreement with those of the halo stars in the same early-type galaxies.

The relative importance of the form of the color-metallicity relation on the one hand, and
structure in the underlying metallicity distribution on the other hand, remains an
unresolved question, at least in the minds of many.
One argument against the importance of nonlinear color-metallicity relations in
extragalactic GC populations comes from the bimodal GC metallicity distribution in the
Milky Way Galaxy (e.g., Zinn 1985).  
However, it is unclear if the GC system of the Milky Way would be recognized as bimodal
to an outside observer, given the necessity of avoiding disk contamination when selecting
GCs in spirals.  For instance, Zepf \& Ashman (1993) show that the metallicity
distribution of the Milky Way halo GCs, the subpopulation that they considered relevant
for comparison with elliptical galaxies, is unimodal.
Furthermore, the generalization of the Milky Way's properties to galaxies such as NGC\,1399,
which has $\sim\,$40 times as many GCs and must have had a much more chaotic interaction
history than a spiral galaxy in a low-density group, is uncertain.  It is clear that the
metallicity peaks in the Milky Way do not correspond with those derived for giant
ellipticals based on linear transformations of optical colors (e.g., Zepf \& Ashman 1993;
Gebhardt \& Kissler-Patig 1999; Peng \etal\ 2006).

Several other large, nearby galaxies possess extensive GC systems that are probably
better analogues to those of giant ellipticals.  The nearest of these is M31, with roughly 500
confirmed GCs (Huxor \etal\ 2011).  Based on high-quality spectroscopic metallicities for
322 M31 GCs (twice as many as in the whole of the Milky Way), Caldwell \etal\ (2011) find that
the metallicity distribution does not evince clear bimodality, ``in strong distinction
with the bimodal Galactic globular distribution.''  Instead, M31 shows a broad peak at
$\feh\approx-1$, with possibly several minor peaks superposed like ripples on the broader
distribution.
NGC\,5128 (Centaurus~A) is another galaxy with an extensive GC system having a broad,
complex spectroscopic GC metallicity distribution with perhaps three closely spaced peaks
(Beasley \etal\ 2008; Woodley \etal\ 2010; see discussion in Yoon \etal\ 2011b).  The
evidence suggests that both of these large nearby galaxies have had more active
formation histories than has the Milky Way, although given their locations in small groups,
it is unlikely that their histories were as turbulent as those of cluster ellipticals.

The recent innovative studies by Foster \etal\ (2010, 2011) find that metallicity
distributions of GCs inferred from the calcium triplet (CaT) index differ from those found by
simple linear transformation of the optical colors.  Foster \etal\ (2011) report that the
peak in metallicity from their CaT measurements in the elliptical galaxy NGC\,4494
actually corresponds to the trough in the $g{-}i$ distribution, and they note the
similarity of this result to their CaT study of the elliptical NGC\,1407 (Foster
\etal\ 2010), as well as the results on M31 from Lick indices by Caldwell \etal\ (2011).
The authors conclude that the lack of bimodality in the CaT distributions, and the
similarity in the CaT-derived metallicities for bright GCs on opposite sides of the color
trough, ``casts serious doubt on the reliability of the CaT as a metallicity indicator''
(Foster \etal\ 2011).  However, they also acknowledge that the very different color and
metallicity distributions in these elliptical galaxies can be reconciled if the optical
color bimodality is mainly the result of nonlinearity in the color-metallicity relation.
In this case, the trough in the color distribution would correspond to a minimum in the
slope of metallicity as a function of color; thus, GCs straddling the color divide
would have nearly the same metallicities, as observed.

Further evidence on both sides of the bimodality debate from spectroscopic and
optical/near-IR observations of extragalactic GC systems has been discussed by Blakeslee
\etal\ (2010a; see Sec.~4 of that work) and in even greater detail by Yoon \etal\ (2011b;
see their Sec.~2).  We refer the reader to those works for more information.  The main
conclusions were that while evidence exists for breaks in the metallicity distributions
of some galaxies, clear bimodality like that seen in the Milky Way is rare.  In
particular, the distribution in metallicity usually appears less bimodal than that of the
optical colors, which suggests that at least some of the observed color bimodality is
likely the result of the color-metallicity relation.  Pipino \etal\ (2007) arrived at a
similar conclusion, based mainly on the spectroscopic results of Puzia \etal\ (2005).
More recently, Alves-Brito \etal\ (2011) have found spectroscopic evidence for bimodality
in the metallicities of GCs in the Sombrero galaxy (NGC\,4594), a nearly edge-on Sa
galaxy with a prominent bulge.  Thus, in contrast to M31 and Cen~A, this may be an
example of another galaxy, in addition to the Milky Way, with a truly bimodal GC
metallicity distribution.

We now summarize the conclusions from this optical-IR study of GCs in NGC\,1399:
\begin{itemize}
\item{The optical \gz, \gI, \VI\ color distributions are bimodal; the evidence is much
    weaker for \IH.  If one is determined to enforce bimodality via double Gaussian
    modeling, the resulting bimodal split (dominated by the blue component) is
    different from that found for the optical colors (dominated by the red).  Thus,
    these color bimodalities cannot both be linear reflections of an underlying
    metallicity bimodality.}
\item{These ``differing bimodalities'' imply that there must be a nonlinear relation
  between the purely optical and mixed optical-IR colors; indeed, we find empirically
  that the dependence of optical-IR \IH\ color on optical \gI\ and \gz\ is
  nonlinear.  The significance of the curvature indicates that it is not simply a
  case of photometric errors obscuring bimodality in \IH.}
\item{At least some of the colors studied here must vary nonlinearly with
  metallicity, with the slope of the metallicity dependence having a local
  extremum somewhere within the observed range of these data.  Observations and
  modeling both indicate that, for the case of GC populations, broad baseline
  optical colors are more likely to exhibit this type of nonlinearity.}
\item{Thus, the ``universal'' bimodality in the optical GC color distributions
  of elliptical galaxies is at least partly the result of 
 the inflected form of the nonlinear color-metallicity relations.}
%
\end{itemize}

Determining the full importance of the nonlinear behavior of GC colors, and the precise
form of the underlying metallicity distributions, will require further effort in amassing
large samples of high-quality optical/near-IR photometry and spectroscopy, as well as in
improved theoretical modeling.  As we have done here, it is essential to compare the
proportions in any proposed bimodal decompositions to check for consistency among
techniques, or across wavebands, rather than simply quoting the significance at which
single-Gaussian models can be rejected.
We are currently analyzing the optical-IR GC color distributions for the full sample of
16 galaxies from our \hst\ WFC3/IR project (H.~Cho \etal, in preparation); this and other
ongoing projects will continue to push back the obscuring veil over extragalactic GC
metallicity distributions
and permit more useful constraints on the formation histories of massive early-type galaxies.

\acknowledgements  
We are grateful to Patrick~C{\^ o}t{\' e}, Suk-Jin Yoon, and Michele Cantiello for many
helpful discussions, and Caroline Foster for helpful email correspondence.  
We thank the anonymous referee for an extremely thorough check of our
analysis and for providing useful comments.
H.C.~acknowledges support from the National Research Foundation of Korea (NRF) 
Global Internship Program and from the NRF to the Center for Galaxy Evolution
Research; she thanks the National Research Council of Canada's Herzberg
Institute of Astro\-physics for hospitality during her visit.
E.W.P.~acknowledges the support of the Peking University Hundred Talent Fund
(985). A.J.~acknowledges support from BASAL CATA PFB-06, FONDAP CFA 15010003,
Ministry of Economy ICM Nucleus P07-021-F and Anillo ACT-086.

{\it Facility:} \facility{HST (WFC3/IR, ACS/WFC)}

 \clearpage
 \begin{deluxetable}{ccccccc}
\tablecaption{Summary of Observing Programs for NGC\,1399\label{tab:obs}}
\tablewidth{0pt}
\tablehead{
\colhead{Instr.\ \&} &
\colhead{Prog} &
\colhead{Dataset} &
\colhead{Bandpass} &
\colhead{Exp.\ time} &
\colhead{$m_1$\tmk{a}} &
\colhead{mag\tmk{b}} \\
\colhead{Channel} &
\colhead{(GO)} &
\colhead{} &
\colhead{} &
\colhead{(sec)} &
\colhead{(AB)} &
\colhead{}
}
\startdata
WFC3/IR & 11712 & \dataset[ADS/Sa.HST#IB1H08020]{IB1H08020}  &  F160W &  1197 & 25.960 & $H_{160}$ \\
ACS/WFC & 10911 & \dataset[ADS/Sa.HST#J9P305010]{J9P305010}  &  F814W &  1224 & 25.937 & $I_{814}$ \\
ACS/WFC & 10911 & \dataset[ADS/Sa.HST#J9P305020]{J9P305020}  &  F475W & \p680 & 26.068 & $g_{475}$ \\
ACS/WFC & 10129 & \dataset[ADS/Sa.HST#J8ZQ01010,ADS/Sa.HST#J8ZQ02010,ADS/Sa.HST#J8ZQ03010,ADS/Sa.HST#J8ZQ04010]{J8ZQ0$[$1--4$]$010} &  F606W &  2108 & 26.486 & $V_{606}$ \\
ACS/WFC & 10217 & \dataset[ADS/Sa.HST#J90X02010]{J90X02010}  & F850LP &  1220 & 24.862 & $z_{850}$ \\
ACS/WFC & 10217 & \dataset[ADS/Sa.HST#J90X02020]{J90X02020}  & F475W  & \p760 & 26.068 & $g_{475}$ 
\tablenotetext{a}{Photometric zero point: the AB magnitude for corresponding to one
  count per second.}
\tablenotetext{b}{The symbol used here to denote the
  AB-calibrated magnitudes and other quantities in the given bandpass.}
\enddata
\end{deluxetable}

\begin{deluxetable}{crrrrrrrrc}
\tablecaption{GMM Results for Matched Samples}
\tablewidth{0pt}
\tablehead{
\colhead{Quant} &
\colhead{$N$} &
\colhead{\phantom{+}min$\,:\,$max} &
\colhead{kurt} &
\colhead{\phantom{<}$p$-val} &
\colhead{$p_1$} &
\colhead{$p_2$} &
\colhead{$D$} &
\colhead{frac(2)} &
\colhead{\textit{bi?}}\\
\colhead{(1)} &
\colhead{(2)} &
\colhead{(3)} &
\colhead{(4)} &
\colhead{(5)} &
\colhead{(6)} &
\colhead{(7)} &
\colhead{(8)} &
\colhead{(9)} &
\colhead{(10)} 
}
\startdata
     $g-I$ & 584 &  $0.36:1.60$ &$-$0.58 & ${<\,}0.001$ &    0.83 $\pm$ 0.02 &    1.19 $\pm$ 0.01 & 3.4 $\pm$ 0.3 &    0.68 $\pm$ 0.04 & Y\phantom{?} \\ 
     $g-I$ & 579 &  $0.64:1.47$ &$-$0.93 & ${<\,}0.001$ &    0.82 $\pm$ 0.01 &    1.18 $\pm$ 0.01 & 3.8 $\pm$ 0.2 &    0.71 $\pm$ 0.03 & Y\phantom{?} \\ 
     $V-I$ & 578 &  $0.25:0.76$ &$-$0.86 & ${<\,}0.001$ &    0.34 $\pm$ 0.01 &    0.54 $\pm$ 0.01 & 3.3 $\pm$ 0.2 &    0.74 $\pm$ 0.03 & Y\phantom{?} \\ 
     $I-H$ & 345 & $-0.22:1.36$ &   0.28 &   0.266  & $-$0.22 $\pm$ 0.36 &    0.69 $\pm$ 0.10 & 5.5 $\pm$ 1.8 &    1.00 $\pm$ 0.37 & N\phantom{?} \\ 
     $I-H$ & 342 &  $0.10:1.25$ &$-$0.43 &   0.110  &    0.67 $\pm$ 0.13 &    0.94 $\pm$ 0.09 & 1.7 $\pm$ 0.4 &    0.08 $\pm$ 0.30 & N\phantom{?} \\ 
     $V-H$ & 341 &  $0.09:1.86$ &$-$0.25 &   0.010  &    1.01 $\pm$ 0.13 &    1.39 $\pm$ 0.10 & 1.6 $\pm$ 0.4 &    0.39 $\pm$ 0.22 & N\phantom{?} \\ 
     $V-H$ & 339 &  $0.27:1.86$ &$-$0.54 &   0.012  &    0.88 $\pm$ 0.18 &    1.34 $\pm$ 0.12 & 2.1 $\pm$ 0.5 &    0.62 $\pm$ 0.29 & Y? 
\enddata
\label{tab:gmm_indsamp}
\tablecomments{Columns list:  (1)~quantity analyzed;
  (2)~number of objects in the analyzed sample;
  (3)~minimum and maximum values of the given quantity for the GCs in
  the sample;
  (4)~kurtosis of the distribution;
  (5)~GMM ``$p$~value,'' indicating the significance of the preference for a double Gaussian over a
  single Gaussian model (lower $p$ values are more significant);
  (6)~mean and uncertainty of the first peak in the double Gaussian model;
  (7)~mean and uncertainty of the second peak in the double Gaussian model;
  (8)~separation $D$ of the peaks in units of the Gaussian~$\sigma$;
  (9)~fraction of GC candidates assigned to the second Gaussian
  component (or to the ``red peak'');
  (10)~assessment of the evidence for bimodality.
}
\end{deluxetable}

 \begin{deluxetable}{rccccccc}
\tablecaption{Matched WFC3/IR and ACS Photometry for GC Candidates}
\tablewidth{0pt}
\tablehead{
\colhead{ID} &
\colhead{R.A.} &
\colhead{Dec.} &
\colhead{\Iacs } &
\colhead{$(\gIacs)_{r=3}$} &
\colhead{$(\IHacs)_{r=3}$} &
\colhead{$(\gIacs)_{\rm cor}$} &
\colhead{$(\IHacs)_{\rm cor}$}
\\
\colhead{~ } &
\colhead{(J2000)} &
\colhead{(J2000)} &
\colhead{(mag)} &
\colhead{(mag)} &
\colhead{(mag)} &
\colhead{(mag)} &
\colhead{(mag)} \\
\colhead{(1)} &
\colhead{(2)} &
\colhead{(3)} &
\colhead{(4)} &
\colhead{(5)} &
\colhead{(6)} &
\colhead{(7)} &
\colhead{(8)} 
}
\startdata
  14 & 54.603067 & $-$35.467057 & 22.892 $\pm$ 0.023 & 1.125 $\pm$ 0.067 & 0.642 $\pm$ 0.047 & 1.145 &  0.622 \\ 
  26 & 54.603686 & $-$35.467382 & 23.390 $\pm$ 0.029 & 1.283 $\pm$ 0.077 & 0.831 $\pm$ 0.062 & 1.303 &  0.811 \\ 
  40 & 54.603018 & $-$35.464966 & 21.741 $\pm$ 0.012 & 1.212 $\pm$ 0.037 & 0.881 $\pm$ 0.020 & 1.232 &  0.861 \\ 
 198 & 54.605421 & $-$35.464718 & 20.974 $\pm$ 0.013 & 1.277 $\pm$ 0.040 & 0.878 $\pm$ 0.018 & 1.297 &  0.858 \\ 
 269 & 54.605974 & $-$35.464424 & 21.457 $\pm$ 0.012 & 0.924 $\pm$ 0.033 & 0.473 $\pm$ 0.021 & 0.944 &  0.453 \\ 
 300 & 54.608004 & $-$35.467911 & 23.063 $\pm$ 0.021 & 1.198 $\pm$ 0.054 & 0.607 $\pm$ 0.041 & 1.218 &  0.587 \\ 
 346 & 54.606921 & $-$35.464733 & 22.779 $\pm$ 0.019 & 1.209 $\pm$ 0.050 & 0.719 $\pm$ 0.034 & 1.229 &  0.699 \\ 
 355 & 54.608060 & $-$35.466963 & 22.781 $\pm$ 0.018 & 1.220 $\pm$ 0.049 & 0.838 $\pm$ 0.031 & 1.240 &  0.818 \\ 
 368 & 54.605945 & $-$35.461861 & 22.172 $\pm$ 0.016 & 0.891 $\pm$ 0.039 & 0.491 $\pm$ 0.030 & 0.911 &  0.471 \\ 
\dots &\dots &\dots &\dots &\dots &\dots &\dots &\dots 
\enddata
\label{tab:data}
\tablecomments{Columns list:  (1)~ID number;
  (2)~right ascension;
  (3)~declination;
  (4)~SExtractor \magauto\ value from F814W image;
  (5)~\gIacs\ color for $r{\,=\,}3$ pix measurement aperture;
  (6)~\IHacs\ color for $r{\,=\,}3$ pix measurement aperture;
  (7)~\gIacs\ color, with estimated correction for differential aperture effects;
  (8)~\IHacs\ color, with estimated correction for differential aperture effects.
All magnitudes and colors are corrected for Galactic extinction.  Note
that the analysis and fits presented in this paper are based on the
aperture colors given in columns~(5) and~(6).\\
(This table is available in its entirety in a machine-readable form in the online journal. A portion is shown here for guidance regarding
its form and content.)
}
\end{deluxetable}

\begin{deluxetable}{crrrrrrrrc}
\tablecaption{GMM Results for Merged GC Sample}
\tablewidth{0pt}
\tablehead{
\colhead{Quant} &
\colhead{$N$} &
\colhead{\phantom{+}min$\,:\,$max} &
\colhead{kurt} &
\colhead{\phantom{<}$p$-val} &
\colhead{$p_1$} &
\colhead{$p_2$} &
\colhead{$D$} &
\colhead{frac(2)} &
\colhead{\textit{bi?}}
\\
\colhead{(1)} &
\colhead{(2)} &
\colhead{(3)} &
\colhead{(4)} &
\colhead{(5)} &
\colhead{(6)} &
\colhead{(7)} &
\colhead{(8)} &
\colhead{(9)} &
\colhead{(10)} 
}
\startdata
     $g-I$ & 322 &  $0.64:1.52$ &$-$0.71 & ${<\,}0.001$ &    0.84 $\pm$ 0.02 &    1.20 $\pm$ 0.01 & 3.6 $\pm$ 0.3 &    0.73 $\pm$ 0.05 & Y\phantom{?} \\ 
     $g-z$ & 322 &  $0.76:1.79$ &$-$0.90 & ${<\,}0.001$ &    0.98 $\pm$ 0.04 &    1.41 $\pm$ 0.03 & 3.2 $\pm$ 0.2 &    0.72 $\pm$ 0.07 & Y\phantom{?} \\ 
     $I-H$ & 322 & $-0.22:1.25$ &   0.24 &   0.052  &    0.66 $\pm$ 0.30 &    0.93 $\pm$ 0.11 & 1.7 $\pm$ 1.4 &    0.10 $\pm$ 0.36 & N\phantom{?} \\ 
     $I-H$ & 320 &  $0.10:1.25$ &$-$0.45 &   0.055  &    0.66 $\pm$ 0.12 &    0.93 $\pm$ 0.08 & 1.7 $\pm$ 0.4 &    0.10 $\pm$ 0.29 & N\phantom{?} \\ 
     $I-H$ & 316 &  $0.21:1.18$ &$-$0.65 &   0.062  &    0.62 $\pm$ 0.11 &    0.92 $\pm$ 0.08 & 2.0 $\pm$ 0.3 &    0.24 $\pm$ 0.28 & N\phantom{?} \\ 
$[$Fe/H$]$ & 298 & $-2.87:0.76$ &   0.78 & ${<\,}0.001$ & $-$1.33 $\pm$ 0.25 & $-$0.23 $\pm$ 0.06 & 2.0 $\pm$ 0.6 &    0.78 $\pm$ 0.08 & N\phantom{?} \\ 
$[$Fe/H$]$ & 285 & $-1.83:0.76$ &$-$0.22 &   0.002  & $-$1.49 $\pm$ 0.18 & $-$0.29 $\pm$ 0.07 & 3.3 $\pm$ 0.5 &    0.92 $\pm$ 0.08 & N\phantom{?} 
\enddata
\label{tab:gmm_mrgsamp}
\tablecomments{Columns are the same as in Table~\ref{tab:gmm_indsamp}.}
\end{deluxetable}

\end{CJK}


\begin{thebibliography}{}
%
\bibitem[Alves-Brito et al.(2011)]{2011MNRAS.tmp.1306A} Alves-Brito, A., 
Hau, G.~K.~T., Forbes, D.~A., et al.\ 2011, \mnras, 417, 1823
%
\bibitem[Ashman et al.(1994)]{1994AJ....108.2348A} Ashman, K.~M., Bird, 
C.~M., \& Zepf, S.~E.\ 1994, \aj, 108, 2348 
%
\bibitem[Ashman 
\& Zepf(1992)]{1992ApJ...384...50A} Ashman, K.~M., \& Zepf, S.~E.\ 1992, \apj, 384, 50 
%
\bibitem[DeGraaff et al.(2007)]{2007ApJ...671.1624D} Barber DeGraaff, R., 
Blakeslee, J.~P., Meurer, G.~R., \& Putman, M.~E.\ 2007, \apj, 671, 1624 
%
\bibitem[Bailin 
\& Harris(2009)]{2009ApJ...695.1082B} Bailin, J., \& Harris, W.~E.\ 2009, \apj, 695, 1082 %
\bibitem[Bassino et al.(2006)]{2006A&A...451..789B} Bassino, L.~P.,
Faifer, F.~R., Forte, J.~C., et al.\ 2006, \aap, 451, 789
%
\bibitem[Beasley et al.(2002)]{2002MNRAS.333..383B} Beasley, M.~A., Baugh, 
C.~M., Forbes, D.~A., Sharples, R.~M., \& Frenk, C.~S.\ 2002, \mnras, 333, 383 
%
\bibitem[Beasley et al.(2008)]{2008MNRAS.386.1443B} Beasley, M.~A., 
Bridges, T., Peng, E., et al.\ 2008, \mnras, 386, 1443 
%
\bibitem[Bergbusch \& VandenBerg(2001)]{2001ApJ...556..322B} Bergbusch, P.~A., \& VandenBerg, D.~A.\ 2001, \apj, 556, 322 
%
\bibitem[Bertin \& Arnouts 1996]{667} Bertin, E. \& Arnouts, S. 1996, \aaps,  117, 393
%
\bibitem[Blakeslee et al.(2003)]{2003ASPC..295..257B} Blakeslee, J.~P.,
Anderson, K.~R., Meurer, G.~R., Ben{\'{\i}}tez, N., \& Magee, D.\ 2003,
Astronomical Data Analysis Software and Systems XII, 295, 257
%
%
%
\bibitem[Blakeslee et al.(2009)]{2009ApJ...694..556B} Blakeslee, J.~P., et 
al.\ 2009, \apj, 694, 556
%
\bibitem[Blakeslee et al.(2010a)]{2010ApJ...710...51B} Blakeslee, J.~P., 
Cantiello, M., \& Peng, E.~W.\ 2010a, \apj, 710, 51 
%
\bibitem[Blakeslee et al.(2010b)]{2010ApJ...724..657B} Blakeslee, J.~P., et 
al.\ 2010b, \apj, 724, 657 
%
\bibitem[Caldwell et al.(2011)]{2011AJ....141...61C} Caldwell, N., 
Schiavon, R., Morrison, H., Rose, J.~A., \& Harding, P.\ 2011, \aj, 141, 61 
%
\bibitem[Cantiello \& Blakeslee(2007)]{2007ApJ...669..982C}
 Cantiello, M., \& Blakeslee, J.~P.\ 2007, \apj, 669, 982 
%
\bibitem[Chies-Santos et al.(2011a)]{2011A&A...525A..20C} Chies-Santos, A.~L.,
Larsen, S.~S., Kuntschner, H., et al.\ 2011a, \aap, 525, A20
%
\bibitem[Chies-Santos et al.(2011b)]{2011A&A...525A..19C} Chies-Santos, A.~L.,
Larsen, S.~S., Wehner, E.~M., et al.\ 2011b, \aap, 525, A19
%
\bibitem[Cote et al.(1998)]{1998ApJ...501..554C} C{\^o}t{\'e}, P., Marzke, R.~O., 
\& West, M.~J.\ 1998, \apj, 501, 554 
\bibitem[C{\^o}t{\'e} et al.(2004)]{2004ApJS..153..223C} C{\^o}t{\'e}, P., 
et al.\ 2004, \apjs, 153, 223 
%
\bibitem[Dirsch et al.(2003)]{2003AJ....125.1908D} Dirsch, B., Richtler, 
T., Geisler, D., Forte, J.~C., Bassino, L.~P., \& Gieren, W.~P.\ 2003, \aj, 125, 1908 
%
\bibitem[Dotter et al.(2007)]{2007AJ....134..376D} Dotter, A., et al.\ 2007, \aj, 134, 376 
%
\bibitem[Dotter et al.(2010)]{2010ApJ...708..698D} Dotter, A., et al.\ 2010, \apj, 708, 698 
%
\bibitem[Dressel et al.(2010)]{wfc3handbook} Dressel, L. \etal\ 2010. Wide
Field Camera~3 Instrument Handbook, Version~3.0 (Baltimore: STScI), 
http://www.stsci.edu/hst/wfc3
%
\bibitem[Drinkwater et al.(2001)]{2001ApJ...548L.139D} Drinkwater, M.~J., 
Gregg, M.~D., \& Colless, M.\ 2001, \apjl, 548, L139 
%
%
\bibitem[Forbes et al.(1997)]{1997AJ....113.1652F} Forbes, D.~A., Brodie, 
J.~P., \& Grillmair, C.~J.\ 1997, \aj, 113, 1652 
%
\bibitem[Forte et al.(2007)]{2007MNRAS.382.1947F} Forte, J.~C., Faifer, F., 
\& Geisler, D.\ 2007, \mnras, 382, 1947 
%
\bibitem[Foster et al.(2010)]{2010AJ....139.1566F} Foster, C., Forbes, 
D.~A., Proctor, R.~N., Strader, J., Brodie, J.~P., 
\& Spitler, L.~R.\ 2010, \aj, 139, 1566 
%
\bibitem[Foster et al.(2011)]{2011MNRAS.415.3393F} Foster, C., et al.\ 2011, \mnras,  415, 3393 
%
\bibitem[Fruchter \& Hook(2002)]{2002PASP..114..144F} Fruchter, A.~S.~\& Hook, R.~N.\ 2002, \pasp, 114, 144 
%
\bibitem[Gebhardt \& Kissler-Patig(1999)]{1999AJ....118.1526G} Gebhardt, K., \& Kissler-Patig, M.\ 1999, \aj, 118, 1526 
%
\bibitem[Hanes \& Harris(1986)]{1986ApJ...309..564H} Hanes, D.~A., \& Harris, W.~E.\ 1986, \apj, 309, 564 
%
\bibitem[Harris(1991)]{1991ARA&A..29..543H} Harris, W.~E.\ 1991, \araa, 29, 543 
\bibitem[Harris(2009)]{2009ApJ...703..939H} Harris, W.~E.\ 2009, \apj, 703, 939 
%
\bibitem[Harris et al.(2006)]{2006ApJ...636...90H} Harris, W.~E., Whitmore, 
B.~C., Karakla, D., Oko{\'n}, W., Baum, W.~A., Hanes, D.~A., 
\& Kavelaars, J.~J.\ 2006, \apj, 636, 90 
%
\bibitem[Hartigan \& Hartain(1985)]{dip} 
Hartigan, J. A., \& Hartigan, P. M. 1985, Ann.\ Stat., 13, 70
%
%
\bibitem[Huxor et al.(2011)]{2011MNRAS.414..770H} Huxor, A.~P., Ferguson, 
A.~M.~N., Tanvir, N.~R., et al.\ 2011, \mnras, 414, 770 
%
\bibitem[Jefferys et al.(1988)]{1988CeMec..41...39J} Jefferys, W.~H., 
Fitzpatrick, M.~J., \& McArthur, B.~E.\ 1988, Celestial Mechanics, 41, 39 
%
\bibitem[Jord{\'a}n et al.(2004)]{2004ApJS..154..509J} Jord{\'a}n, A., \etal\  2004, \apjs, 154, 509 
\bibitem[Jord{\'a}n et al.(2005)]{2005ApJ...634.1002J} Jord{\'a}n, A., 
 et al.\ 2005, \apj, 634, 1002 
\bibitem[Jord{\'a}n et al.(2007)]{2007ApJS..169..213J} Jord{\'a}n, A., \etal\  2007, \apjs, 169, 213 
%
\bibitem[Jord{\'a}n et al.(2009)]{2009ApJS..180...54J} Jord{\'a}n, A., et 
al.\ 2009, \apjs, 180, 54 
%
\bibitem[Kalirai et al.(2009)]{2009wfc..rept...30K} Kalirai, J.~S., 
MacKenty, J., Bohlin, R., Brown, T., Deustua, S., Kimble, R.~A., 
\& Riess, A.\ 2009, Instrument Science Report WFC3 2009-30
%
\bibitem[King(1966)]{1966AJ.....71...64K} King, I.~R.\ 1966, \aj, 71, 64 
%
%
\bibitem[Kissler-Patig et al.(2002)]{2002A&A...391..441K} Kissler-Patig, M.,
  Brodie, J.~P., \& Minniti, D.\ 2002, \aap, 391, 441 
%
\bibitem[Kissler-Patig et al.(1998a)]{1998AJ....115..105K} Kissler-Patig, M., et al.\ 1998a, \aj, 115, 105 
%
\bibitem[Kissler-Patig et al.(1998b)]{1998MNRAS.298.1123K} Kissler-Patig, 
M., Forbes, D.~A., \& Minniti, D.\ 1998b, \mnras, 298, 1123 
%
\bibitem[Koekemoer et~al. 2002]{692} Koekemoer, A.M., Fruchter, A.S., Hook,
R.N., \& Hack, W. 2002, in The 2002 HST Calibration Workshop: Hubble after the
Installation of the ACS and the NICMOS Cooling System, 337
%
\bibitem[Kotulla et al.(2008)]{2008MNRAS.387.1149K} Kotulla, R., Fritze, U., \&
Anders, P.\ 2008, \mnras, 387, 1149
%
\bibitem[Kravtsov \& Gnedin(2005)]{2005ApJ...623..650K} Kravtsov, A.~V., \& Gnedin,
O.~Y.\ 2005, \apj, 623, 650
%
\bibitem[Kundu \& Zepf(2007)]{2007ApJ...660L.109K} Kundu, A., \& Zepf, S.~E.\ 2007,
\apjl, 660, L109
%
%
%
\bibitem[Lee et al.(2002)]{2002AJ....124.2664L} Lee, H.-c., Lee, Y.-W., 
\& Gibson, B.~K.\ 2002, \aj, 124, 2664 
%
\bibitem[Lee et al.(1994)]{1994ApJ...423..248L} Lee, Y.-W., Demarque, P., 
\& Zinn, R.\ 1994, \apj, 423, 248 
%
\bibitem[Liu et al.(2011)]{2011ApJ...728..116L} Liu, C., Peng, E.~W., 
Jord{\'a}n, A., Ferrarese, L., Blakeslee, J.~P., C{\^o}t{\'e}, P., 
\& Mei, S.\ 2011, \apj, 728, 116 
%
\bibitem[Masters et al.(2010)]{2010ApJ...715.1419M} Masters, K.~L., et al.\ 
2010, \apj, 715, 1419 
%
\bibitem[Mei et al.(2005)]{2005ApJ...625..121M} Mei, S., et al.\ 2005, \apj, 625, 121 
%
\bibitem[Mieske et al.(2006)]{2006ApJ...653..193M} Mieske, S., et al.\ 2006, \apj, 653, 193 
\bibitem[Mieske et al.(2010)]{2010ApJ...710.1672M} Mieske, S., et al.\ 2010, \apj, 710, 1672 
%
\bibitem[Muratov \& Gnedin(2010)]{2010ApJ...718.1266M} Muratov, A.~L.,
  \& Gnedin, O.~Y.\ 2010, \apj, 718, 1266  
\bibitem[Ostrov et al.(1998)]{1998AJ....116.2854O} Ostrov, P.~G., Forte, 
J.~C., \& Geisler, D.\ 1998, \aj, 116, 2854 
\bibitem[Ostrov et al.(1993)]{1993AJ....105.1762O} Ostrov, P., Geisler, D., 
\& Forte, J.~C.\ 1993, \aj, 105, 1762 
%
\bibitem[Peng et al.(2006)]{2006ApJ...639...95P} Peng, E.~W., et al.\ 2006, \apj, 639, 95 
\bibitem[Peng et al.(2009)]{2009ApJ...703...42P} Peng, E.~W., et al.\ 2009, \apj, 703, 42 
%
\bibitem[Pipino, Puzia, \& Matteucci(2007)]{pipino07} Pipino, A., Puzia, T. H., \& Matteucci, F., 2007, \apj, 665, 295
%
\bibitem[Pirzkal et al.(2011)]{2011wfc..rept...11P} Pirzkal, N., Mack, J., 
Dahlen, T., \& Sabbi, E.\ 2011, STScI Instrument Science Report ISR WFC3-2011-11
%
\bibitem[Puzia et al.(2002)]{2002A&A...391..453P} Puzia, T.~H., Zepf, S.~E., Kissler-Patig, M., et al.\ 2002, \aap, 391, 453 
%
\bibitem[Puzia et al.(2005)]{2005A&A...439..997P} Puzia, T.~H., Kissler-Patig, M., Thomas, D., Maraston, C., Saglia, R.~P., Bender, R., Goudfrooij, P., \& Hempel, M.\ 2005, \aap, 439, 997 
%
%
\bibitem[Richtler(2006)]{2006BASI...34...83R} Richtler, T.\ 2006, Bulletin 
of the Astronomical Society of India, 34, 83 
%
\bibitem[Sarajedini et al.(1997)]{1997PASP..109.1321S} Sarajedini, A., Chaboyer, B., \& Demarque, P.\ 1997, \pasp, 109, 1321 
\bibitem[Schlegel et~al. 1998]{708} Schlegel, D.J., Finkbeiner, D.P., 
\& Davis, M. 1998, \apj, 500, 525
%
\bibitem[Schuberth et al.(2010)]{2010A&A...513A..52S} Schuberth, Y., Richtler, T., Hilker, M., et al.\ 2010, \aap, 513, A52 
%
\bibitem[Sirianni et~al. 2005]{sirianni+05} Sirianni, M., et~al. 2005, \pasp, 
117, 1049
%
\bibitem[Strader et al.(2006)]{2006AJ....132.2333S} Strader, J., Brodie, 
J.~P., Spitler, L., \& Beasley, M.~A.\ 2006, \aj, 132, 2333 
%
\bibitem[Tonry et~al. 1997]{716} Tonry, J.~L., Blakeslee, J.~P., Ajhar,
  E.~A., \& Dressler, A., 1997, \apj, 475, 399
\bibitem[Villegas et al.(2010)]{2010ApJ...717..603V} Villegas, D., et al.\ 
2010, \apj, 717, 603 
%
%
%
\bibitem[Woodley et al.(2010)]{2010ApJ...708.1335W} Woodley, K.~A., Harris, 
W.~E., Puzia, T.~H., et al.\ 2010, \apj, 708, 1335 
%
\bibitem[Yi et al.(2001)]{2001ApJS..136..417Y} Yi, S., Demarque, P., Kim, 
Y.-C., et al.\ 2001, \apjs, 136, 417 
%
\bibitem[Yoon et al.(2006)]{2006Sci...311.1129Y} Yoon, S.-J., Yi, S.~K., 
\& Lee, Y.-W.\ 2006, Science, 311, 1129 
%
\bibitem[Yoon et al.(2011a)]{Yoon2011a}Yoon, S.-J., \etal\ 2011a, \apj, 743, 149 (arXiv:1109.5174)
\bibitem[Yoon et al.(2011b)]{Yoon2011b}Yoon, S.-J., \etal\ 2011b, \apj, 743, 150 (arXiv:1109.5178)
\bibitem[Zepf \& Ashman(1993)]{1993MNRAS.264..611Z} Zepf, S.~E., \& Ashman, K.~M.\ 1993, \mnras, 264, 611 
\bibitem[Zinn(1985)]{1985ApJ...293..424Z} Zinn, R.\ 1985, \apj, 293, 424 
%
\end{thebibliography}
\end{document}